\documentclass[12pt, draftclsnofoot, onecolumn]{IEEEtran}

\usepackage{mymacros}
\usepackage{tcolorbox}
\usepackage{soul}

\usepackage{multirow}
\newcolumntype{P}[1]{>{\centering\arraybackslash}p{#1}}
\newcolumntype{M}[1]{>{\centering\arraybackslash}m{#1}}
\usepackage{url}
\usepackage{setspace}

\begin{document}

\title{Indoor Millimeter-Wave Systems: Design and Performance Evaluation}
%
%
%

\author{Jacek Kibi\l{}da, Allen B. MacKenzie, Mohammad J. Abdel-Rahman, Seong Ki Yoo, Lorenzo Galati Giordano, Simon L. Cotton, Nicola Marchetti, Walid Saad, William G. Scanlon, Adrian Garcia-Rodriguez, David L\'opez-P\'erez, Holger Claussen, and Luiz A. DaSilva
\thanks{J. Kibi\l{}da, N. Marchetti, and L. A. DaSilva are with CONNECT, Trinity College, The University of Dublin, Ireland, E-mails: \{kibildj,nicola.marchetti,dasilval\}@tcd.ie.}
\thanks{L. A. DaSilva is also with the Commonwealth Cyber Initiative (CCI) and Electrical and Computer Engineering, Virginia Tech, US, E-mail: ldasilva@vt.edu.}
\thanks{A.~B.~MacKenzie, M. J. Abdel-Rahman, and W.~Saad are with Electrical and Computer Engineering, Virginia Tech, US, E-mails: \{mackenab,mo7ammad,walids\}@vt.edu.}
\thanks{A.~B.~MacKenzie is also with the Electrical and Computer Engineering Department at Tennessee Tech, US, E-mail: amackenzie@tntech.edu.}
\thanks{M. J. Abdel-Rahman is also with the Electrical Engineering and Computer Science Departments at Al Hussein Technical University, Jordan.}
\thanks{S. L. Cotton is with Centre for Wireless Innovation, ECIT Institute, Queen's University Belfast, UK, E-mail: simon.cotton@qub.ac.uk}
\thanks{S. K. Yoo is with School of Computing, Electronics and Mathematics, Coventry University, UK, E-mail: ad3869@coventry.ac.uk}
\thanks{W. G. Scanlon is with Tyndall National Institute, Ireland, E-mail: w.scanlon@ieee.org.}
\thanks{L. Galati Giordano, A. Garcia-Rodriguez, D. L\'opez-P\'erez, and H. Claussen are with Nokia Bell Labs, Ireland, E-mails: \{lorenzo.galati\_giordano,holger.claussen\}@nokia-bell-labs.com,a.garciarodriguez.2013@ieee.org,dr.david.lopez@ieee.org}
}

%
%

\markboth{Proceedings of the IEEE}%
{Kibi\l{}da \MakeLowercase{\textit{et al.}}: Indoor Millimeter-Wave Systems}
%



\maketitle

\begin{abstract}
Indoor areas, such as offices and shopping malls, are a natural environment for initial \ac{mmWave} deployments. While we already have the technology that enables us to realize indoor \ac{mmWave} deployments, there are many remaining challenges associated with system-level design and planning for such. The objective of this article is to bring together multiple strands of research to provide a comprehensive and integrated framework for the design and performance evaluation of indoor \ac{mmWave} systems. The paper introduces the framework with a status update on \ac{mmWave} technology, including ongoing \ac{5G} wireless standardization efforts, and then moves on to experimentally-validated channel models that inform performance evaluation and deployment planning. Together these yield insights on indoor \ac{mmWave} deployment strategies and system configurations, from feasible deployment densities to beam management strategies and necessary capacity extensions.

\end{abstract}


%
\IEEEpeerreviewmaketitle

\vspace*{-0.2in}

\section{Introduction}
\label{sec:intro}

\IEEEPARstart{O}{ur} motivation in this article is to present a comprehensive framework for performance evaluation and design practices dedicated to indoor \acf{mmWave} networks. The framework integrates inputs from multiple areas of \ac{mmWave} networking expertise, from standardization, to channel measurements and modelling, to system-level evaluations and deployment planning. Its major contribution is insights into indoor \ac{mmWave} deployment planning strategies and system configurations that are grounded and informed by experimentally-validated channel models.

\acp{mmWave} are a natural choice for mobile indoor deployments due to much shorter link distances, weak penetration through walls, and large available bandwidths. In fact, the concept of using \acp{mmWave} to provision multi-Gbps speeds over indoor areas dates back to works published in the late 90's \cite{OhmoriYamaoNakajima_2000,Takimoto_1997}. Later came a suite of standards that offered the possibility of ad hoc networking, including ECMA-387 \cite{seyedi2009physical}, IEEE 802.15.3c \cite{BaykasSumLanEtAl_2011}, and more recently IEEE Wireless Gigabit 802.11ad/ay \cite{6979964_80211ad_magazine,8088544_80211ay_magazine}. Nowadays, the importance of \acp{mmWave} keeps increasing, since they are one of the key mobile broadband networking features of the \acf{5G} wireless \cite{5GNRUnveilingTheEssential_Ericsson2018}, \cite{dahlman20185g}.

\begin{figure*}[!t]
\centering
\includegraphics[width=6.5in]{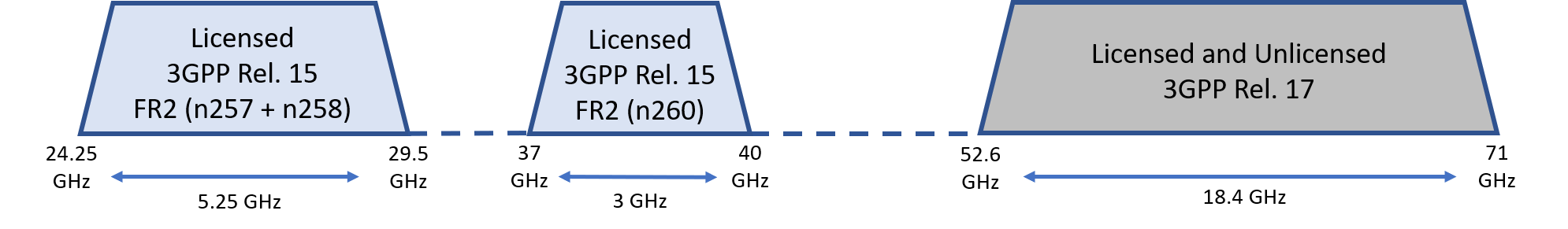}
\caption{Status of the \acl{mmWave} spectrum for \ac{5G}-\ac{NR}.}
\label{fig:mmWaveSpectrum}
\end{figure*}

At the time of writing, the first set of \ac{5G}-\ac{NR} specifications had already been defined as part of the \ac{3GPP} Release 15, functionally frozen in September 2018. Work on Release 16 is ongoing and is planned to be frozen in June 2020, with Release 17 to follow. The \ac{FR} considered for deploying \ac{mmWave} technology, as agreed in Release 15, is from \unit[24.25]{GHz} to \unit[52.6]{GHz}, also referred to as \ac{FR}2\footnote{\ac{FR}1 addresses the spectrum between \unit[410]{MHz} and \unit[7.125]{GHz}.}. \ac{FR}2 currently addresses three bands: \unit[24.25--27.5]{GHz} (n257), \unit[26.5--29.5]{GHz} (n258), and \unit[37--40]{GHz} (n260), all meant to support \ac{TDD} operation only. Usage of frequencies between \unit[52.6]{GHz} and \unit[71]{GHz} is currently under study in Release 17. In particular, the spectrum  around \unit[60]{GHz} presents an attractive use case, as it: a) does not require licenses, and b) is harmonised globally \cite{5GNRU_Giupponi2018}. In this spectrum, the \ac{3GPP} is planning to deploy \ac{NR}-based access technology -- \ac{NR-U} -- which will incorporate extensions necessary to work in unlicensed spectrum. \ac{NR-U} will account for national restrictions of the \unit[60]{GHz} band, which the \ac{3GPP} had already identified in Release 14 \cite{3GPP38805}. In \Fig{mmWaveSpectrum}, we summarize the adoption of \ac{mmWave} spectrum for \ac{5G}-\ac{NR}.

As part of its \ac{5G} activity, the \ac{3GPP} defined the \ac{eMBB} Indoor Hotspot scenario as a natural case study for indoor \ac{mmWave} mobile networks \cite{3gpp38.913_2018}. The key purpose of this scenario is to provision small-cell coverage of high data rates and capacity to a user population within a confined area. Typical examples for this type of scenario include an open office, airport hall, or shopping mall. For such a setup, one would be interested in understanding the required density of access points, deployment locations, and effective network settings. Understanding these system design aspects requires adequate system-level evaluation that involves similarly adequate representation of the \ac{mmWave} channel.

There are many ongoing measurement campaigns being conducted around the globe with the aim of characterizing and modeling the \ac{mmWave} propagation channel, e.g., \cite{MiWEBA1,mmMagic1,MaccartneyRappaportSunEtAl_2015,HurBaekKimEtAl_2016,YooUE2eNB}. 
By their nature, all these measurements are context-specific (as any experimental work), and, depending on the modeling technique used, pertain to different types of system-level evaluations. For example, in an open office scenario, radio infrastructure is typically mounted to the ceiling or walls, illuminating the main area inside \cite{ITU-RM2412_2017}. In such a setup, the main factor limiting signal propagation is shadowing by physical objects \cite{NiuLiJinEtAl_2015}. In particular, human bodies introduce extra attenuation, referred to as body blockage
, which may vary with the orientation and position of the bodies with respect to both the device and the serving access point \cite{KiYooCottonJinChunEtAl_2017}. However, whether the communication link is blocked or not will also have a discernible effect on the fading characteristics of the associated wireless channel. Broadly speaking, blocked links often display a richer multipath structure, with weaker direct component and larger delay spread, than the unobstructed ones \cite{MaccartneyRappaportSunEtAl_2015}. Moreover, it was found that within indoor locations, such as large offices and hallways, body-related blockage has a less pronounced effect on the received signal power, most likely due to the increased scattering in the environment \cite{KiYooCottonJinChunEtAl_2017}. The trick is to capture all of these propagation-related effects in a model that is both accurate and amenable to the analysis at system-level.

In what follows we describe our framework by first discussing our channel measurement campaign and a modelling approach that we subsequently integrate into both system-level evaluations and deployment planning. Using this framework, we provide insights on feasible deployment densities, beam management strategies, and necessary capacity extensions. While the framework can be applied to any indoor \ac{mmWave} networking scenario, our case study focuses on an open office environment for illustrative purposes.

\section{Indoor \ac{mmWave} channel}
\label{sec:channel}
It is well known that many aspects of wireless communications, e.g., system design, network topology and performance, are dependent upon an accurate understanding of the channel characteristics. Therefore, channel characteristics at \ac{mmWave} must be comprehensively studied to allow detailed channel models to be developed. This knowledge will both inform the design of future \ac{mmWave} communications systems and help predict important performance measures such as the achievable coverage probabilities and capacities. In wireless communications channels, the characteristics of the received signal are often characterized in terms of path loss, shadowing and small-scale fading. 
In what follows, we introduce the $\kappa$-$\mu$ fading model and provide empirical evidence for its utility and versatility in the context of \ac{mmWave} communications, in particular for indoor hotspot \ac{eMBB} applications, which is one of the five test  use cases selected in \cite{ITU-RM2412_2017}.

\subsection{Indoor \ac{mmWave} Channel Measurements}

A number of studies have recently been conducted in \cite{3GPP2, mmMagic1, 6979964_80211ad_magazine, IEEE802_1, METIS, RappaportMacCartneySamimiEtAl_2015, Vitucci_2019}, which are key to the provisioning of future \ac{5G} services at \ac{mmWave} frequencies for both indoor and outdoor environments. 
In most previous \ac{mmWave} channel studies \cite{METIS, RappaportMacCartneySamimiEtAl_2015, Vitucci_2019}, the measurements have considered the case where both the \ac{TX} and \ac{RX} are stationary and mounted on stands. In some of these studies \cite{METIS, RappaportMacCartneySamimiEtAl_2015}, obstructions in the optical \ac{LOS} path between the \ac{TX} and \ac{RX} are analyzed. Crucially, however, none of these studies considers scenarios where the TX and/or RX are in close proximity to the human user body. To this end, in this work, we have considered the case where a hypothetical \ac{UE} (i.e., the TX in our case) is held by the user while emulating different \ac{UE} operating modes. Furthermore, the user is moving through the environment. 
This is driven by the need to understand the potential impact of \ac{UE} operation mode and user mobility on the small-scale fading characteristics, knowledge of which will be essential for determining the localized performance of future user-centric \ac{mmWave} networks.   

\subsubsection{Measurement System and Environments}
\begin{figure*}[!t]
\centering
\includegraphics[width=5.5in]{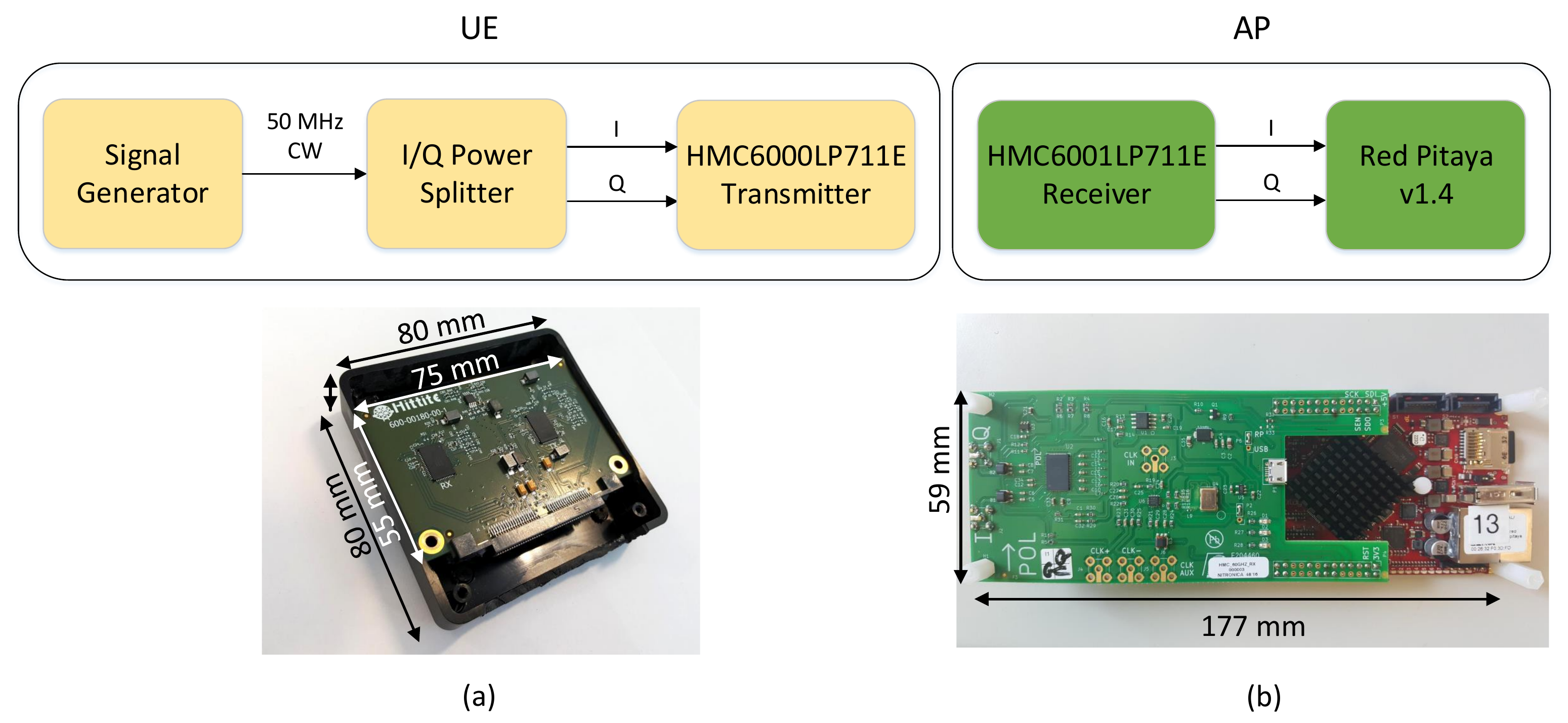}
\caption{Hypothetical (a) \ac{UE} and (b) \ac{AP} used for the \ac{mmWave} channel measurements.}
\label{fig:UE_AP}
\end{figure*}

\begin{figure*}[!t]
\centering
\includegraphics[width=5.5in]{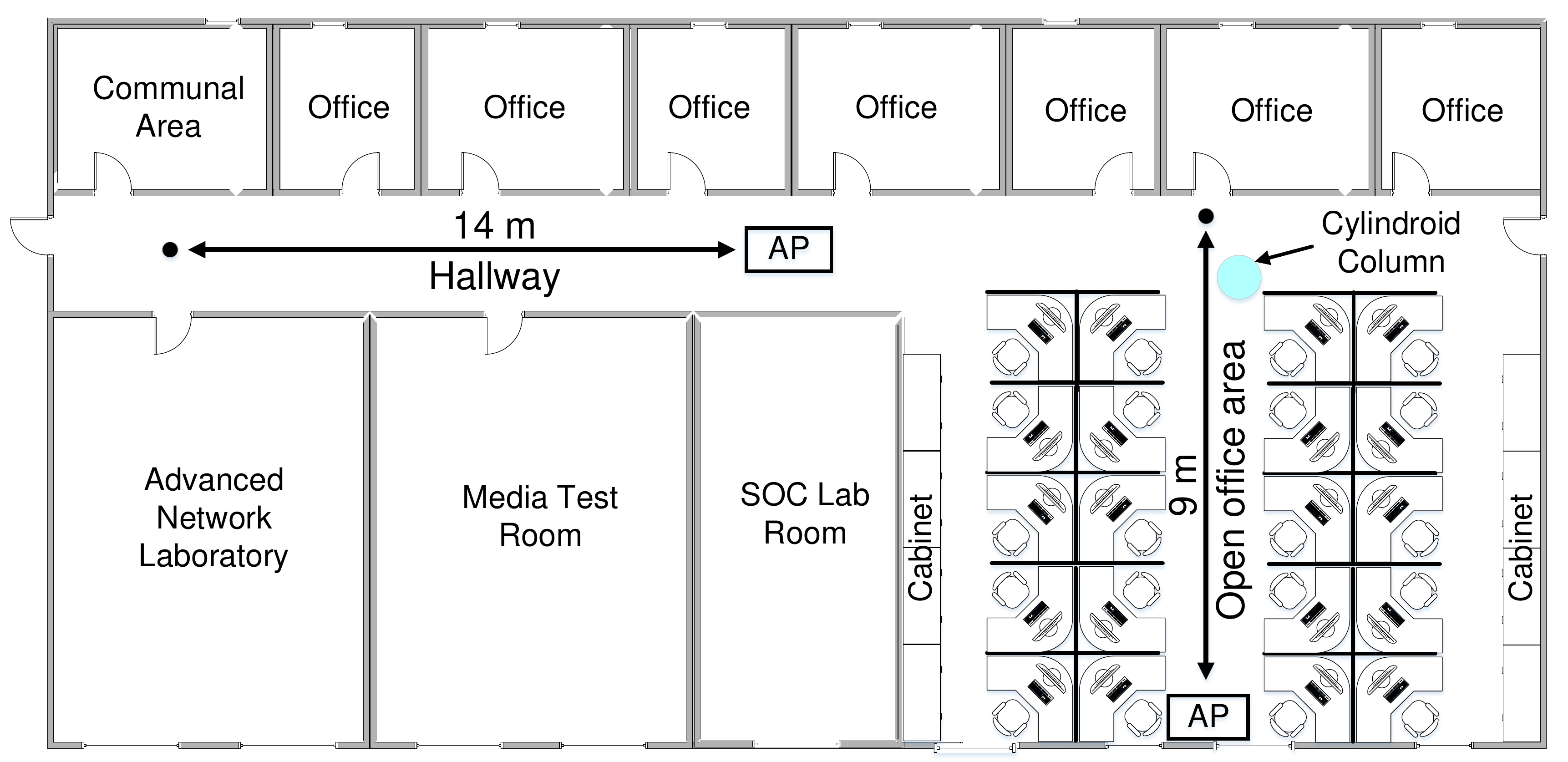}
\caption{Indoor hallway and open office environments considered in this study along with annotated user trajectories.}
\label{fig:Environment}
\end{figure*}

To emulate a possible indoor hotspot \ac{eMBB} use case the \ac{mmWave} channel between a \ac{UE} and \ac{AP} is considered. The hypothetical \ac{UE} and \ac{AP} used for the \ac{mmWave} channel measurements are shown in \Fig{UE_AP}(a) and (b), respectively. Details of the hypothetical \ac{UE} and \ac{AP} can be found in \cite{YooEuCAP2018}. The channel measurements were conducted within a hallway (\unit[17.38]{m} $\times$ \unit[1.40]{m}) and an open office area (\unit[10.62]{m} $\times$ \unit[12.23]{m}) as shown in \Fig{Environment}. Both the hallway and open office environments are located on the 1\textsuperscript{st}  floor of the Institute of Electronics, Communications and Information Technology (ECIT) at Queen's University Belfast in the United Kingdom. Both environments featured metal studded dry walls with a metal tiled floor covered with polypropylene-fiber, rubber backed carpet tiles, and metal ceiling with mineral fiber tiles and recessed louvered luminaries suspended \unit[2.70]{m} above floor level. The open office area contained a number of soft partitions, cabinets, personal computers, chairs and desks. It should be noted that both the hallway and open office environments were unoccupied for the duration of the channel measurements and the \ac{AP} was placed above a ceiling tile with the antenna boresight oriented towards the floor, i.e., imitating a ceiling-mounted \ac{AP}.  

\subsubsection{Usage Cases}
During the measurements, the \ac{UE} was operated by an adult male of \unit[1.83]{m} in height and mass \unit[80]{kg}. A number of different use cases likely to be encountered in everyday scenarios were considered. These were: (1) operating an app, where the user held the \ac{UE} with his two hands in front of his body; (2) carrying a device I, where the \ac{UE} was located in the right-front pocket of the user's clothing; (3) carrying a device II, where the user held the \ac{UE} with his right hand beside his right leg. Herein, and for brevity, we denote the three different \ac{UE} usage cases as \emph{\app}, \emph{\hand} and \emph{\pocket} respectively. In this study, we considered two different dynamic channel conditions, namely, (1) \emph{mobile \ac{LOS}} and (2) \emph{mobile \ac{NLOS}}, where the user walked towards and away from the hypothetical \ac{AP} in a straight line, respectively. It is worth remarking that the \ac{NLOS} channel conditions occurred when the direct optical path between the \ac{UE} and \ac{AP} was obstructed by the user's body, i.e., self-blockage.

To improve the validity and robustness of the parameter estimates obtained in this study, all the measurements were repeated three times. Due to the dissimilar sizes of each environment, the considered walking distances were different. In particular, these were \unit[9]{m} and \unit[14]{m} for the open office area and hallway respectively. The average walking speed maintained by the user throughout all of the experiments was approximately \unit[1]{m/s}.

\subsubsection{Path Loss}

The path loss is a measure of the signal attenuation between the transmitter and receiver as a function of the separation distance and can be expressed as \cite{pathloss}
\begin{equation} \label{eq:pathloss}
P\left[{\rm dB}\right] = P_{0}\left[{\rm dB}\right] + 10 \alpha \log_{10} \left({d}/{d_0}\right),
\end{equation}
\noindent 
where $P_{0}$ represents the path loss at the reference distance $d_{0}$, $\alpha$ is the path loss exponent which indicates the rate at which the path loss increases with distance and $d$ is the separation distance between the transmitter and receiver. To obtain estimates for $P_{0}$ and $\alpha$, we first removed the \ac{EIRP} and gain at the receiver from the raw signal power received by the \ac{AP} and then performed linear regression. To enable this, the elapsed time was first converted into a distance using the test user's velocity. The reference distance, which should be in far field region of the antenna, was chosen to be \unit[1]{m} for all environments. The mean values of the parameter estimates for $P_{0}$ and $\alpha$ averaged over all the trials for all of the considered cases are given in \Tab{para_estimate} along with the body blockage, which is defined as the difference between the path loss at the reference distance (i.e., $P_0$) for the \ac{LOS} and \ac{NLOS} conditions\footnote{Although we define the body blockage as the difference between the path loss at the reference distance  for the \ac{LOS} and \ac{NLOS} conditions in this paper, the human body blockage loss is also presented in \cite{Reviewer2_ref2} using the double knife-edge diffraction.}.  The path loss exponents for the LOS scenarios were found to be smaller than that associated with isotropic radiation in free space ($\alpha = 2$). This was possibly due to the waveguide effect which can often be present within indoor environments \cite{Reviewer2_ref1}. As anticipated, for both the hallway and office environments, the $P_0$ values for the \ac{NLOS} were greater than those for the \ac{LOS} due to the shadowing effects caused by the test user's body. When considering the values of the body blockage, it was observed that the \pocket case had a smaller body blockage compared to those for the \app and \hand cases. This confirms the intuition that the \ac{UE} to ceiling-mounted \ac{AP} channel is less susceptible to body blockage when the user is carrying the \ac{UE} further away from the body, which is the case with the hand scenario. 

\subsubsection{Small-Scale Fading}
The $\kappa$-$\mu$ distribution has been proposed as a generalized statistical model, which may be used to characterize the random variation of the received signal caused by multipath fading \cite{Yacoub2007}. The \ac{PDF} of the signal envelope, $H$, in a $\kappa$-$\mu$ fading channel can be expressed as 
\begin{equation}\label{eq:kappamu}
f_{{H}}({h}) = \frac{{2\mu {{\left( {\kappa  + 1} \right)}^{\frac{{\mu  + 1}}{2}}}{h}^\mu }}{{{\kappa ^{\frac{{\mu  - 1}}{2}}}\exp \left( {\mu \kappa } \right){\Omega ^{\frac{{\mu  + 1}}{2}}}}}\exp \left( {\frac{{ - \mu \left( {\kappa  + 1} \right){h}^2}}{\Omega }} \right) {I_{\mu  - 1}}\left( {2\mu \sqrt {\kappa \left( {\kappa  + 1} \right)} \frac{{{h}}}{{\sqrt \Omega  }}} \right),
\end{equation}
\noindent where $I_{v}(\cdot)$ represents the modified Bessel function of the first kind with order $v$. In terms of its physical interpretation, $\kappa$ is defined as the ratio between the total power in the dominant signal components and the total power in the scattered signal components, $\mu$ is the number of multipath clusters and $\Omega$ is the mean signal power given by $\Omega = \Eb[H^2]$. 

\begin{table*}[!t]
\footnotesize
\renewcommand{\arraystretch}{1.4}
\caption{Average Parameter Estimates for the Path Loss Model and $\kappa$-$\mu$ Fading Model for All of the Cases}
\vspace{-0.3cm}
\label{tab:para_estimate}
\centering
\begin{tabular}{ |c|c|c|c|c|c|c|c|c|c|c|c|c| }
\hline

{\textbf{Environ-}} & \multirow{2}{*}{\textbf{ Use Case }} &  \multicolumn{5}{c|}{\textbf{\centering{LOS}}} &  \multicolumn{5}{c|}{\textbf{\centering{NLOS}}} & {\textbf{{Body}}} \\

\cline{3-12}
{\textbf{ment}}&  &  \centering{$\bar{\alpha}$} & {$\bar{P_{0}}{\rm [dB]}$} &  \centering{$\bar{\kappa}$} & \centering{$\bar{\mu}$} & \centering{$\bar{\Omega}$} &    \centering{$\bar{\alpha}$} & {$\bar{P_{0}}{\rm [dB]}$} &  \centering{$\bar{\kappa}$} & \centering{$\bar{\mu}$} & {$\bar{\Omega}$} & {\textbf{Blockage}} \\					
					
\hline

\multirow{3}{*}{\textbf{Hallway}} & \app & 1.92 & 78.31 & 2.80 &  0.77 & 1.16 &   1.93 & 95.39 & 0.67 & 0.96 & 1.25 & 17.09 dB\\
\cline{2-13}
 &\hand & 1.92 & 82.55 & 2.64 & 0.78 & 1.17 &  1.95 & 95.60 & 0.47 & 1.02 & 1.24 & 13.05 dB\\
\cline{2-13}
 & \pocket  & 1.93 & 90.42 & 1.89 & 0.88 & 1.18 & 1.94 & 97.49 & 0.89 & 0.99 & 1.22 & 7.06  dB\\
\hline

\multirow{3}{*}{\textbf{Office}}  & \app & 2.58 & 81.31 & 1.14 & 1.00 & 1.21 &  1.03 & 101.41 & 0.48 & 1.00 & 1.26 & 20.09 dB\\
\cline{2-13}
 & \hand & 1.38 & 92.32 & 1.46 & 0.91 & 1.21 &  1.01 & 102.11 & 0.46 & 1.00 & 1.26 & 9.79 dB\\
\cline{2-13}
 & \pocket  & 1.52 & 95.74 & 1.24 & 0.93 & 1.21 &  1.38 & 101.83 & 0.50 & 1.04 & 1.24 & 6.09 dB\\

\hline
\end{tabular}
\end{table*}

The small-scale fading was extracted by removing \emph{both} the path loss and large-scale fading\footnote{The large-scale fading was extracted from the received signal power by first removing the estimated path loss using the parameters given in \Tab{para_estimate}. The resulting dataset was then averaged using a moving window of 100 channel samples (equivalent to a distance of 10 wavelengths).} from the channel data before transforming the data to its linear amplitude. The parameter estimates for the $\kappa$-$\mu$ fading model were obtained using a non-linear least squares routine. \Tab{para_estimate} provides the mean parameter estimates for the $\kappa$-$\mu$ fading model averaged over all three trials for each of the \ac{UE} to AP channels.

\begin{figure*}[!t]
\centering
\includegraphics[width=6.5in]{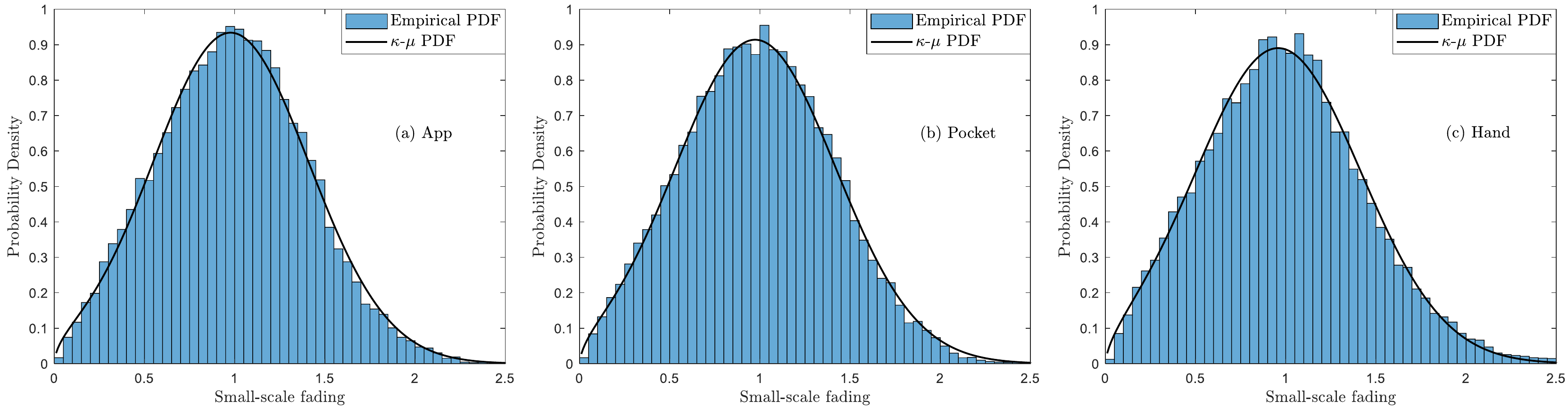}
\caption{Empirical (bars) and theoretical (continuous lines) \acp{PDF} of the small-scale fading observed in the hallway environment for the \ac{LOS} scenario with (a) app, (b) pocket and (c) hand cases.}
\label{fig:Fitting}
\end{figure*}

Our first observation is that the obtained results provide evidence for the correctness of our methodology. For all of the \ac{LOS} scenarios and environments, the $\kappa$ parameters were found to be greater than unity ($\kappa > 1$), indicating the presence of a strong dominant signal component. In contrast, for the \ac{NLOS} scenarios (i.e., when the direct signal path was blocked by user's body), the $\kappa$ parameters were smaller than unity ($\kappa < 1$), indicating the absence of a dominant signal contribution. Additionally, for the \ac{NLOS} scenarios, the $\kappa$ parameters obtained for the \pocket case were slightly greater than those for the \app and \hand cases, suggesting that the \ac{UE} to ceiling-mounted \ac{AP} link for the \pocket is less impacted by the human body blockage. The $\mu$ parameters for the \ac{LOS} scenarios were slightly smaller than those obtained for the \ac{NLOS} scenarios. This suggests that the signal fluctuation observed in the \ac{LOS} scenarios is less impacted by multipath clustering compared to that experienced in \ac{NLOS}. As an example of the fits obtained, \Fig{Fitting} presents the empirical \acp{PDF} of the small-scale fading alongside the $\kappa$-$\mu$ \acp{PDF} for all of the \ac{UE} usage cases while the operator walked towards the hypothetical \ac{AP} in the hallway environment. It is clear that the $\kappa$-$\mu$ fading model is in excellent agreement with the measurement data.

To ascertain the most probable model between the $\kappa$-$\mu$ and Rayleigh distributions for characterizing the small-scale fading observed in the \ac{UE}-to-\ac{AP} channels, the \ac{AIC} was employed. More specifically, the \ac{AICc} was used in a similar manner to the analysis employed in \cite{yoo2017composite}, such that
\begin{equation}\label{eq:AIC}
\mathrm{AIC}_c =  - 2\ln \left( {l\left( {\theta |\text{data}} \right)} \right) + 2M + \frac{{2M\left( {M + 1} \right)}}{{n - M - 1}}
\end{equation}
\noindent where $\ln \left( {l\left( {\theta |\text{data}} \right)} \right)$ is the value of the maximized log-likelihood for the unknown parameter $\theta$ of the model given the data, $M$ is the number of estimated parameters available in the model, and $n$ is the sample size. It should be noted that the lower the \ac{AICc} value, the more likely the candidate model was to have generated the data. As shown in Table~\ref{tab:AIC}, the $\kappa$-$\mu$ and Rayleigh fading models were ranked according to their \ac{AICc}. It is clear that the $\kappa$-$\mu$ distribution was selected as the best model for all of the considered cases, suggesting that the added complexity (i.e., additional parameters) in the $\kappa$-$\mu$ model is worthwhile.

\begin{table*}[!t]
\small
\renewcommand{\arraystretch}{1.3}
  \centering
  \vspace{-0.3cm}
  \caption{$AIC_c$ rank for all of the considered UE-to-AP channel measurement data.}
    \begin{tabular}{|c|c|c|c|c|c|}
    \hline
     \multirow{2}{*}{Environment} & \multirow{2}{*}{Use Case} &  \multicolumn{2}{c|}{\textbf{\centering{LOS}}} & \multicolumn{2}{c|}{\textbf{\centering{NLOS}}} \\
     \cline{3-6}
     &  &  $\kappa$-$\mu$  & {Rayleigh} &   $\kappa$-$\mu$  &  {Rayleigh}  \\
    \hline
     \multirow{3}{*}{Hallway} & app & \textbf{1}  & 2 & \textbf{1}  & 2 \\
     \cline{2-6}
      & pocket & \textbf{1}  & 2 & \textbf{1}  & 2 \\
      \cline{2-6}
       & hand & \textbf{1}  & 2 & \textbf{1}  & 2 \\
    \hline
     \multirow{3}{*}{Office} & app & \textbf{1}  & 2 & \textbf{1}  & 2 \\
     \cline{2-6}
      & pocket & \textbf{1}  & 2 & \textbf{1}  & 2 \\
      \cline{2-6}
       & hand & \textbf{1}  & 2 & \textbf{1}  & 2 \\   
    \hline
    \end{tabular}%
  \label{tab:AIC}%
\end{table*}%

\section{System-Level Modelling and Performance Evaluation}
\label{sec:system_level}
Indoor \ac{mmWave} networks will be characterised by much shorter distances between access points and users as well as deployments that will be confined to closed areas and made independent from outdoor deployments due to weak out-in penetration \cite{RanganRappaportErkip_2014}. In \Sec{channel}, we also reported that in the presence of a human user body the indoor \ac{mmWave} channel becomes highly complex with a mix of large- and small-scale fading effects. Understanding how the basic network parameters such as access point density, antenna beamwidth, and serving distance will impact network performance in these conditions requires new and extensive system-level modelling approaches. 

In this section we develop an analytical framework and quantify the performance of indoor \ac{mmWave} networks. Our framework bridges the statistical channel models obtained from measurement campaigns, described in \Sec{channel}, with stochastic network geometry, which reflects the fact that at any given point in time the set of user locations and/or transmitter directionalities may be considered random. Our framework yields analytical expressions describing the distribution of the \ac{SINR} in a generic indoor \ac{mmWave} network. While the obtained expressions are too complex to be amenable to intuition, they allow us to generate numerical results that we can use to cross-validate corresponding Monte Carlo simulations. Ultimately, we use both the analytical expressions and simulations to explore the relationship between basic network parameters, channel scenarios and network performance indicators: coverage, \ac{ATC}, and \ac{EDR}, which we define in \Sec{performance_indicators}.

Most work to date on system-level evaluations for \ac{mmWave} networking are focused on large-scale outdoor areas, e.g., \cite{AkdenizLiuSamimiEtAl_2014,bai2015coverage,TurgutGursoy_2017}. Indoor area \ac{mmWave} network analysis has received much less attention, with some earlier works considering device-to-device applications \cite{venugopal2015interference,VenugopalValentiHeath_2016}, and more recent works addressing hotspot deployments \cite{FiryagunaKibildaGaliottoEtAl_2017,niknam2018interference,azimi2019stochastic,FiryagunaKibildaGaliottoEtAl_2019}. While building on the relevant literature, our work presented in this section goes beyond it in two important ways. First, expanding on our work in \cite{KibildaChunFiryagunaEtAl_2018}, we offer analysis based on the experimentally-validated $\kappa$-$\mu$ fading model, which allows us to consider the impact that different device usage scenarios have on network performance. Then, the network setup we consider is aligned with network setups recommended for system-level evaluation of mobile \ac{mmWave} systems \cite{ITUR_2083,3gpp38.913_2018}. For these we provide analytical expressions describing network performance which, for special cases, can be reduced to closed-form expressions.

\subsection{System Model}

\subsubsection*{Network Geometry}

We consider a network with $n_{\tx}$ transmitters $\Phi=\{x_0,x_1,\ldots,x_{n_{\tx}-1}\}\subset\Wc$, whose locations are uniformly distributed over a finite arbitrary area $\Wc\subset\Rs$. Our analysis takes the perspective of the \emph{reference receiver}, which is an arbitrary point in $\Wc$ at distance $\rho_0$ from the center of mass of $\Wc$. Since we are interested in analyzing feasible networks, we consider $\Wc$ to always contain the reference receiver and the transmitter serving that reference receiver (the \emph{serving transmitter}), which, in our notation, is always indexed as 0. On the horizontal plane, the reference receiver and the serving transmitter are a distance $r_0\geq 0$ apart. We will refer to the wireless link between the reference receiver and its transmitter as the \emph{reference link}. Transmissions from remaining $n_{\tx}^\prime= n_{\tx}-1$ transmitters will be treated as interference to the reference link. We assume that all transmitters are located at height $h_{\tx}$ above the ground, with the reference receiver operating at height $h_{\rx}$. 

\subsubsection*{Directionality Gain}

All transmitters and receivers utilize directional transmission and reception. We assume the transmitter (receiver) antenna gain is calculated based on the cone-bulb model \cite{FiryagunaKibildaGaliottoEtAl_2017}. In this model the main-lobe gain $\gainM_{\tx}$ is a function of the beamwidth $\omega_{\tx}$, normalized over the whole spherical surface:
\begin{equation}
\gainM_{\tx} \frac{1 - \cos(\omega_{\tx}/2)}{2} - \gainm_{\tx}\frac{1 + \cos(\omega_{\tx}/2)}{2}=1,
\label{eq:mainlobe_gain}
\end{equation}
where $\gainm_{\tx}$ is the side-lobe gain. Correspondingly, we will use $\gainM_{\rx}$, $\gainm_{\rx}$, and $\omega_{\rx}$ to refer to the main-lobe gain, side-lobe gain, and beamwidth of the receiver antenna. Usage of the cone-bulb model and the main-lobe beamwidth as our parameter abstracts our analysis from the choice of the antenna system, and makes our results applicable to scenarios with fixed directional antennas. This could be the case, for example, in lecture halls or open offices where users remain static. Conversion between beamwidth, antenna gain, and number of antenna elements for various types of antennas can be found in \cite{Balanis_2012}. Let us note that the cone-bulb model is also a risk-averse choice as it leads to underestimation of coverage results \cite{YuZhangHaenggiEtAl_2017}. 

Then the directionality gain is the product of transmitter and receiver antenna gains, which depend on the beam alignment between the two \cite{AndrewsBaiKulkarniEtAl_2017}. Using the convention that the alignment gain between the receiver and a transmitter is random, we can represent it as a four-dimensional categorical random variable $G_i$, with $i\in\{0,\ldots,n_{\tx}-1\}$. This random variable maps from the space of all possible alignments between the transmitter $i$ and the reference receiver to the space of alignment gains $\Gc$ that is formed as a product of the transmitter and receiver antenna gains, with the corresponding probability mass function $p_{g_i} = \Pb(G_i=g_i)$. In practice the alignment gain from the serving transmitter is maintained stable by beam management operations, which corresponds to conditioning $G_0=g_0$, where $G_0$ is the alignment gain on the reference link and $g_0$ a constant.

\subsubsection*{Blockage}

In our framework, link blockage arises from self-blockage, i.e., blockage of the \ac{LOS} signal path between the receiver and the transmitter by the human user. The severity of this blockage will depend on the usage scenarios defined in \Sec{channel}: \emph{\app} with a device held in front of the user, \emph{\hand} with a device held in the front pocket, \emph{\pocket} with a device carried in the hand. For each scenario, we consider fixed blockage on the reference link\footnote{This pertains to the worst/best case analysis.} and probabilistic blockage on all other links. Consequently, blockage is a Bernoulli random variable, with the success probability $p_{\los}$ and $p_{\nlos}=1-p_{\los}$, where the latter is referred to as the blockage probability. Here we opt for a model where $p_{\los}$ (or, indeed, $p_{\nlos}$) is simply a system parameter that reflects the frequency of link blockages. Furthermore, we assume that each transmission link experiences blockages independently of all other links. Numerical evaluations in \cite{FiryagunaKibildaGaliottoEtAl_2019} show that this independence assumption results in negligible differences to the blockage probability.

\subsubsection*{Propagation Model}

We use the model as proposed in \Sec{channel}, but to capture the bifurcation of its parameters (due to link blockage) we adopt the following notation for the path loss and power fading:
\begin{equation}
l_t(r_i) = \gamma_{t}r_i^{\alpha_t}, \qquad
f_{H_{i,t}}(h) = \frac{\theta_{1,t}^{(\mu_t+1)/2}}{\theta_{2,t}^{(\mu_t-1)/2}}h^{(\mu_t-1)/2}\exp\left(-\theta_{1,t} h-\theta_{2,t}\right)I_{\mu_t-1}\left(2\sqrt{\theta_{1,t}\theta_{2,t} h}\right),
\label{eq:app1_kappa_mu}
\end{equation}
where $r_i$, with $i\in\{0,\ldots,n_{\tx}-1\}$, is the distance to the $i$-th transmitter, $t\in\{\los,\nlos\}$ blockage state, $\gamma_{i,t}$ path loss at the reference distance (in linear scale), $\alpha_t$ path loss exponent, $\theta_{1,t}=\frac{\mu_t(1+\kappa_t)}{\sqrt{\Omega_t}}$, and $\theta_{2,t}=\mu_t\kappa_t$. Scattering and diffraction via static objects, such as inner walls or office furniture, are not explicitly modelled, yet their impact is included in the parameter values given in \Tab{para_estimate}. 

\subsubsection*{Signal-to-Interference-and-Noise Ratio}

Given a transmitter at $x_i$ and blockage state $t\in\{\los,\nlos\}$, the instantaneous power received at the reference receiver located in $y_0\in\Wc$ is:
\begin{equation}
S_{i,t} = G_{i}H_{i,t}l_{t}\left(r_i\right),
\label{eq:rx_power}
\end{equation}
where $r_i=\sqrt{||x_i-y_0||^2 + (h_{\tx}-h_{\rx})^2}$ is the distance between the reference receiver and transmitter $i$, $H_{i,t}$ is the blockage-dependent power fading, and $l_{t}\left(r_i\right)$ is the blockage-dependent path attenuation. Then the \ac{SINR}, under blockage state $t$ on the serving channel is given by
\begin{equation}
\sinr = \frac{S_{0,t}}{I + \tau^{-1}},
\label{eq:sinr_general}
\end{equation}
where $S_{0,t}$ is the instantaneous power received from the serving transmitter, $I=\sum_{i=1}^{n_{\tx}^\prime} S_i$ is the interference power, with $S_i=\Eb_{V_i}[S_{i,V_i}]$ being the received signal power from transmitter $i$ averaged over random blockage states $V_i\in\{\los,\nlos\}$ on link $i$, and $\tau$ is the signal-noise ratio\footnote{For simplicity, we assume the transmit power is identical across all transmitters.}.

\subsection{Network Performance Characterization} 
\label{sec:performance_indicators}

We utilize three key performance indicators recommended by the \ac{3GPP} \cite{3gpp38.913_2018}: coverage, \ac{ATC}, and \ac{EDR}. Broadly speaking, the first represents the coverage achievable in our network, the second provides us with information on the expected user throughput, and the third represents the throughput achievable by a subset of users with the worst channel.

\subsubsection*{Coverage} 

We represent the coverage as the probability that the \ac{SINR} is greater than some pre-defined coverage threshold $\zeta$, which is equivalent to finding the \ac{CCDF} of the \ac{SINR}:
\begin{equation}
\Pb(\sinr>\zeta) =  F^c_{\sinr}(\zeta).
\end{equation}

\subsubsection*{Area traffic capacity} 

\ac{ATC} is a measure of the total traffic a network can serve per unit area (in \unit[]{$\text{bit/s/m}^2$}), which can be calculated as:
\begin{equation}
C_{\mathrm{area}} = \lambda \times \mathrm{bw} \times \mathrm{SE},
\label{eq:atc_def}
\end{equation}
where $\lambda$ is the transmitter density, $\mathrm{bw}$ is the transmission bandwidth, and $\mathrm{SE}$ is the spectral efficiency. The spectral efficiency is the average data rate per unit of spectrum and per cell, which can be expressed\footnote{Noticing that the spectral efficiency is an expectation over a positive random variable $\log(1+\sinr)$.} in terms of the \ac{CCDF} of the \ac{SINR} as:
\begin{equation}
\mathrm{SE} = \int_{0}^{\infty} \Pb\left(\log(1+\sinr)>r\right) \dd{r}= \int_{0}^{\infty} \mathrm{F}^c_{\sinr}\left(2^r-1\right) \dd{r}.
\label{eq:spec_eff_def}
\end{equation}

\subsubsection*{Experienced data rate} 

We define the \ac{EDR} as the 5\textsuperscript{th} percentile of the user throughput distribution, which can also be found numerically following the approach proposed in \cite{KibildaDeVeciana_2018}:
\begin{equation}
q_R(\beta=0.05) = \argmin_{u} \Big\{u + \frac{1}{1-\beta}\int_u^{\infty}\mathrm{F}^c_{\sinr}\left(2^{r/\mathrm{bw}}-1\right)\dd{r}\Big\}.
\label{eq:squantile_gen_int}
\end{equation}

\subsection{Analytical Results}

In the following we provide our main analytical results. We start with the distribution of the received power in \Lemma{recv_pwr_statistics}, which we subsequently use to characterize the \ac{CCDF} of the \ac{SINR} for a \ac{mmWave} network distributed over an arbitrary finite area in \Lemma{ccdf_bpp_network}. Finally, we specialize this result to the case of a network distributed over a disk in \Corollary{ccdf_bpp_network_disc}. 

\begin{lemma}[Received Power Statistics]
Given transmitter $i$ at distance $r_i$, directionality gain $g_i$, and blockage state $t$, the \ac{CCDF} of $S_{i,t}$ can be expressed as:
\begin{equation}
	\begin{split}
	\Pb\left(S_{i,t} > x \right) &=\exp(-\theta_{2,t})\sum_{l=0}^{\infty}\frac{\theta_{2,t}^{l}}{l!} \sum_{n=0}^{l+\mu_t-1} \frac{1}{n!}\left( \frac{\theta_{1,t} x}{\vartheta_{i,t}} \right)^n \exp\left( -\frac{\theta_{1,t} }{\vartheta_{i,t}}x\right),
	\end{split}
	\label{eq:recv_pwr_statistics}
\end{equation}
where $\theta_{1,t}$ and $\vartheta_{i,t}$ are provided in \Eq{app1_kappa_mu} and \Eq{app1_001}, respectively.
\label{lem:recv_pwr_statistics}
\end{lemma}

\begin{IEEEproof}
The above formula relies on a simple re-formulation of the \ac{PDF} in \Eq{app1_kappa_mu}, and the subsequent calculation of the resulting \ac{CDF}, which is presented in \App{kappamu_statistics}.
\end{IEEEproof}

\begin{remark}[Other representation of \Lemma{recv_pwr_statistics}] In \Lemma{recv_pwr_statistics} we define the received power distribution using infinite series representation to facilitate further derivations of the \ac{SINR}. However, the distribution in question can also be represented in closed-form, as shown in \cite{Yacoub2007}, utilizing the generalized Marcum-Q function.
\end{remark}

\begin{lemma}[Distribution of SINR for an Arbitrary Area]
Given the distance to the serving transmitter $r_0$, its directionality gain $g_0$ and \ac{LOS} blockage state $t$, the conditional \ac{CCDF} of the \ac{SINR} is given by 
\begin{equation}
	\begin{split}
	&\Pb\left(\sinr > \zeta\big|G_0=g_0, R_0 = r_0, T_0=t\right)=\\
		    & \exp\left(-\theta_{2,t}-\frac{\zeta\theta_{1,t}}{\vartheta_{0,t}\tau}\right)\sum_{l=0}^{\infty}\frac{\theta_{2,t}^{l}}{l!}\sum_{n=0}^{l+\mu_t-1} \frac{1 }{n!} \left(\frac{\zeta\theta_{1,t}}{\vartheta_{0,t}\tau}\right)^n  \sum_{k=0}^n {n\choose k}\tau^{k}\\
		    &\sum_{k_1+k_2+\ldots+k_{n_{\tx}^\prime}=k}{k\choose k_1,k_2,\ldots,k_{n_{\tx}^\prime}}\prod_{1\leq i\leq n_{\tx}^\prime}\sum_{v\in\{\los,\nlos\}} p_{v}\sum_{g\in \Gc} p_{g} (\mu_v)_{k_i} \theta_{1,v}^{\mu_v}g^{k_i}\exp(-\theta_{2,v})\Zf_{k_i},\\
			&\text{where}\quad \Zf_{k_i} = \Eb_R\left[\frac{  l_v^{k_i}(R)}{\left(\frac{\zeta\theta_{1,t}}{\vartheta_{0,t}}g l_v(R) + \theta_{1,v} \right)^{k_i+\mu_v}}\confhypergeom{k_i+\mu_v}{\mu_v}{\frac{\theta_{1,v}\theta_{2,v}}{\frac{\zeta\theta_{1,t}}{\vartheta_{0,t}}g l_v(R) + \theta_{1,v}}}\right],
	\end{split}
	\label{eq:ccdf_bpp_network}
\end{equation}
and $\vartheta_{0,t}=g_{0}l_{t}\left(r_0\right)$, and $\confhypergeom{}{}{}$ is the confluent hypergeometric function.
\label{lem:ccdf_bpp_network}
\end{lemma}
\begin{IEEEproof}
The proof is provided in \App{sinr_deterministic_network}.
\end{IEEEproof}

In our numerical evaluations we will consider the case of $\Wc$ being a disk of fixed radius $\rad$ around the origin. In this case the expectation in \Eq{ccdf_bpp_network} is taken with respect to the distance between the reference receiver and a uniform point located within $\Wc$ with the \ac{PDF} \cite{khalid2013distance}:
\begin{equation}
    f_R(r) =\frac{1}{\pi \rad^2}\begin{cases}2\pi r, \qquad 0\leq r\leq \rad - \rho_0\\
    2 r \arccos{\frac{r^2+\rho_0^2-\rad^2}{2\rho_0r}}, \qquad \rad - \rho_0 \leq r\leq \rad + \rho_0.\\ \end{cases}
\end{equation}
In the case when the reference receiver is at the center of $\Wc$, we get the closed-form result stated in \Corollary{ccdf_bpp_network_disc}.

\begin{remark}[Usability of \Lemma{ccdf_bpp_network}] While the formula in \Eq{ccdf_bpp_network} yields no immediate intuition on \ac{mmWave} network designs, it can be used to numerically evaluate indoor \ac{mmWave} networks under a variety of scenarios corresponding to:
\begin{itemize}
    \item Arbitrarily-sized areas\footnote{Distance distributions corresponding to a variety of useful geometric shapes can be found in \cite{guo2014outage}.} with arbitrary reference receiver locations (to test for potential boundary effects).
    \item Arbitrary path loss exponents, as given in \Tab{para_estimate}.
    \item Other random blockage or antenna gain models.
    \item Deterministic deployments, in which case the expectation in \Eq{ccdf_bpp_network} should be replaced with a summation over a set of pre-defined distances.
\end{itemize}
\end{remark}

\begin{corollary}[Specializing \Lemma{ccdf_bpp_network} to the case of a receiver located in the center of a disk]
Given the distance to the serving transmitter $r_0$, its directionality gain $g_0$ and \ac{LOS} blockage state $t$, we get that $\Zf_{k_i}$ in \Lemma{ccdf_bpp_network} can be represented using the following closed-form expression:
\begin{equation}
	\begin{split}
	&\Zf_{k_i}=\\
	&\frac{2\gamma_v^{k_i}y^{2-k_i\alpha_v}}{\rad^2\theta_{1,v}^{k_i+\mu_v}(2-k_i\alpha_v)}\humbertpsione{k_i+\mu_v}{k_i-2/\alpha_v}{\mu_v}{k_i-2/\alpha_v+1}{\theta_{2,v}}{-\frac{\zeta\theta_{1,t}}{\vartheta_{0,t}\theta_{1,v}}g \gamma_vy^{-\alpha_v}}\Big|_h^{y=\sqrt{\rho^2+h^2}},
	\end{split}
	\label{eq:ccdf_bpp_network_disc}
\end{equation}
where $\Psi_1$ is the Humbert series, which can also be denoted using the Appell series notation as $\appeltwo{a}{b}{-}{c}{c^\prime}{x}{y}$. 
\label{cor:ccdf_bpp_network_disc}
\end{corollary}

\begin{IEEEproof}
The proof is provided in \App{sinr_deterministic_network_disc}.
\end{IEEEproof}

\begin{remark}[Computationally-Efficient Evaluation of \Lemma{ccdf_bpp_network}]
Due to the exponentially growing number of combinations that involve products of hypergeometric series, the expression in \Lemma{ccdf_bpp_network} even for a relatively small number of transmitters may become cumbersome to evaluate numerically. In order to speed up the numerical evaluations one can pre-compute the indices and the individual terms of the multinomial expansion, and store them into two separate matrices. During the evaluation of the expression in \Lemma{ccdf_bpp_network} the elements of the indices matrix can be used to address the elements of the terms matrix, similarly to the approach proposed in \cite{torrieri2012outage}. Finally, calculating expectation from the definition in \Eq{spec_eff_def} quickly becomes computationally challenging, in which case one may derive analytical representation to this expectation using the approach proposed in \cite{ChunCottonDhillonEtAl_2017b}.
\end{remark}

\subsection{Numerical Evaluations}

Here we numerically evaluate system-level implications of indoor \ac{mmWave} deployments, under a variety of body blockage scenarios as described in \Sec{channel}. In our evaluations, we set the parameters for network geometry and configuration following the Indoor Hotspot \ac{eMBB} performance evaluation setup recommended by the \ac{3GPP} and \ac{ITU-R}\footnote{The recommendations follow each other closely, with the ITU-R document offering more detailed assumptions, e.g., about the ceiling or human user heights assumed.}, in \cite{3gpp38.913_2018} and \cite{ITU-RM2412_2017} respectively. For the propagation and channel settings, we consider the models and values coming from measurements presented in \Sec{channel}. We illustrate coverage results based on numerical evaluations of the obtained analytical formulas and Monte Carlo simulations \footnote{Generally speaking, we will use solid lines to plot numerical values coming from analytical expressions and marks to plot the ones coming from Monte Carlo simulations.}, while the \ac{ATC} and \ac{EDR} results come from Monte Carlo simulations only due to high computational footprint of the averaging formulas in \Eq{spec_eff_def} and \Eq{squantile_gen_int}.

\subsubsection{Evaluation Setup}

\begin{figure}
\centering
\includegraphics[width=0.4\columnwidth]{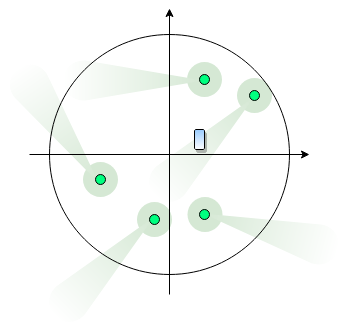}
\caption{Snapshot of the indoor \ac{mmWave} hotspot layout with five transmitters. The reference receiver location is marked with a rectangle and the transmitter locations are the small circles. The direction of transmissions for all the transmitters is illustrated as rotation of the 2D- projection of the cone bulb antenna gain model in \Eq{mainlobe_gain}.}
\label{fig:indoor_hotspot_layout}
\end{figure}

The considered area has a shape of a disk of radius $\rho=\text{\unit[12]{m}}$, which roughly corresponds to the size of the area considered by the \ac{3GPP}. There are $n_{\tx}^\prime$ interfering transmitters, located independently and uniformly within the area, and one transmitter that is also the serving transmitter located a distance $r_0$ apart from the reference receiver in random direction. An example realization of our setup is illustrated in \Fig{indoor_hotspot_layout}.

The deployment is laid out over one floor. According to our measurement setup in \Sec{channel} each transmitter is at a height of $h_{\tx}=\text{\unit[3]{m}}$, while the reference receiver is at a height $h_{\rx}=\text{\unit[1.5]{m}}$. Effectively the reference receiver always stays at a distance greater than the reference distance of \unit[1]{m} from the transmitter. We assume that directionalities of beams produced by $n_{\tx}^\prime$ interfering transmitters are uniform and independent across space, and the transmitter/receiver antenna side-lobe gain is $\gainm_{\tx}=\gainm_{\rx}=\text{\unit[-25]{dBm}}$, while the main-lobe gain is calculated, for a given beamwidth, using \Eq{mainlobe_gain}. The receiver beamwidth is $\omega_{\rx}=30\degree$ and, unless otherwise stated, $n_{\tx}=12$, $\omega_{\tx}=30\degree$, and $r_0=\unit[1]{m}$. We assume that our network uses \unit[200]{MHz} of bandwidth (corresponding to the bandwidth of a component carrier in 5G \cite{ParkvallDahlmanFuruskarEtAl_2017}) for full buffer, downlink transmissions, at a transmit power of \unit[23]{dBm} and the noise figure of \unit[7]{dB}.

Assuming channel reciprocity\footnote{As stated in the Introduction, so far \ac{5G} \ac{NR} \ac{mmWave} technology has been standardized to operate in \ac{TDD} mode only \cite{3gpp38.104_2018}.}, we use the parameter values given in \Tab{para_estimate} while rounding the $\mu$ parameters to the nearest integer value. We take this step to improve the analytical tractability of our derivations and results (see \App{sinr_deterministic_network}) meaning that they can be readily incorporated into other communications performance analyses beyond this study. We would like to highlight that this rounding of the $\mu$ parameter has negligible effect on the observed network performance and is more amenable to the physical interpretation of the results, as the $\mu$ parameter quantifies the number of clusters of scattered multipath. 

In the following we will first look at the network design implications coming from our model and consider the impact that the number of transmitters and their antenna directionality have on the system performance. We perform our numerical analysis for an idealistic case, assuming that our serving transmitter is at a horizontal distance of \unit[1]{m} away from the reference receiver (which, as we will show later, allows us to meet \ac{ATC} and \ac{EDR} targets for most of the analysed scenarios), with its main beam being fully aligned with the receiver beam, and an \ac{LOS} channel between the two. Subsequently, we will re-visit these assumptions and consider how network performance changes with increased distance to the serving transmitter and when the reference link is in \ac{NLOS} state.

\subsubsection{Network performance}

\begin{figure*}[tb!]
\centering
\subfigure[ATC\label{fig:ntx_atc}]{
 \includegraphics[width=0.31\textwidth]{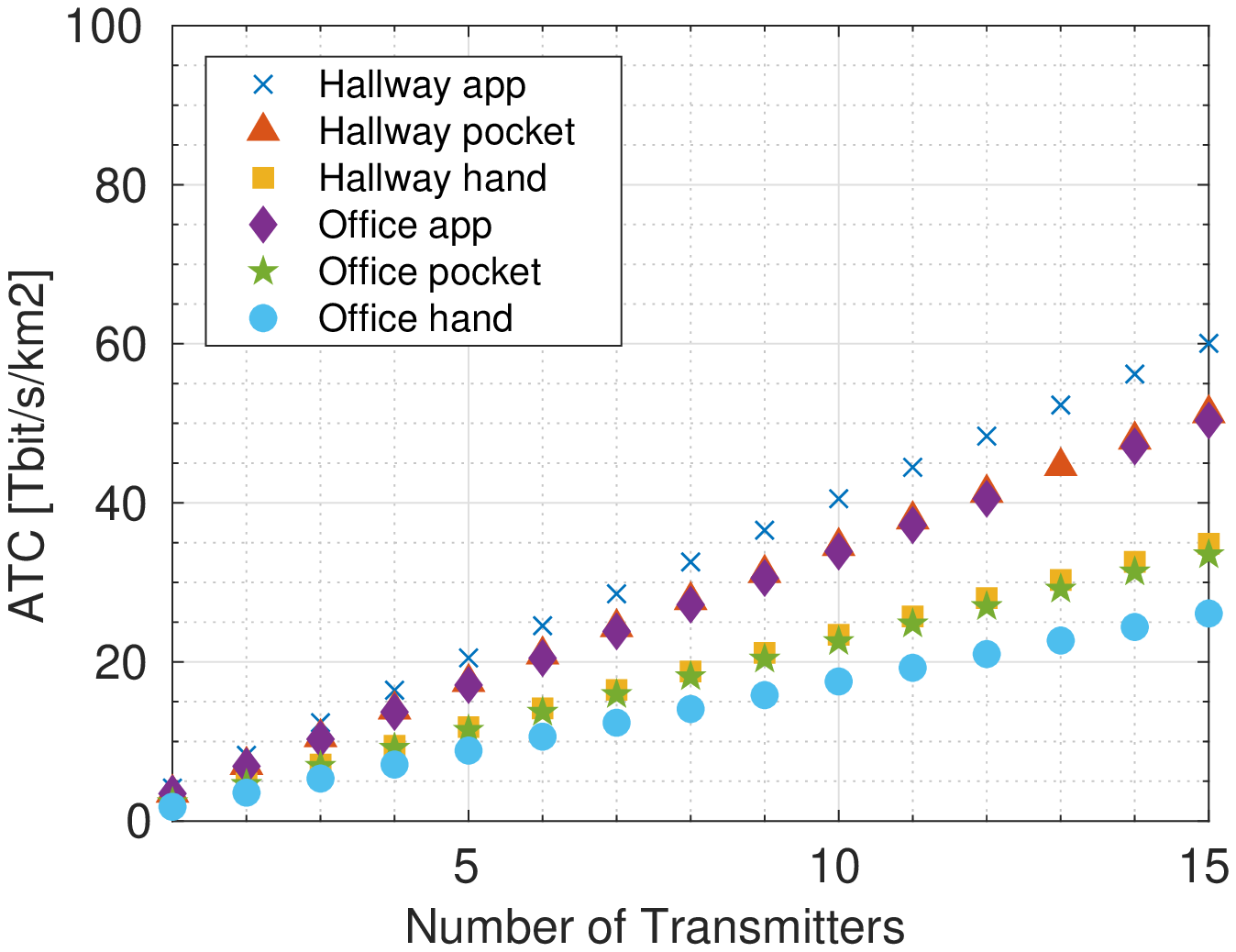}
}
\subfigure[EDR\label{fig:ntx_edr}]{
 \includegraphics[width=0.31\textwidth]{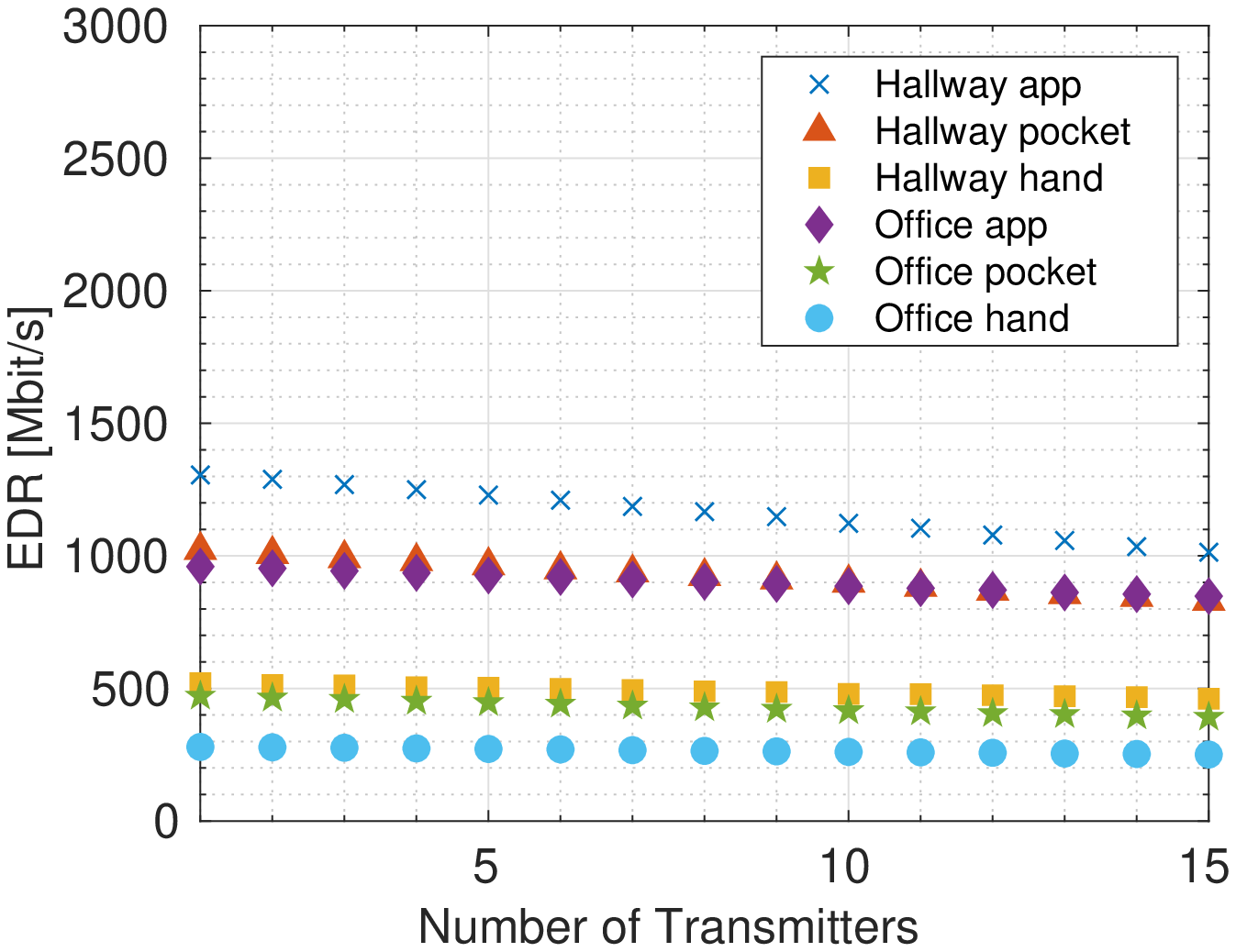}
}
\subfigure[Coverage\label{fig:ntx_cov}]{
 \includegraphics[width=0.31\textwidth]{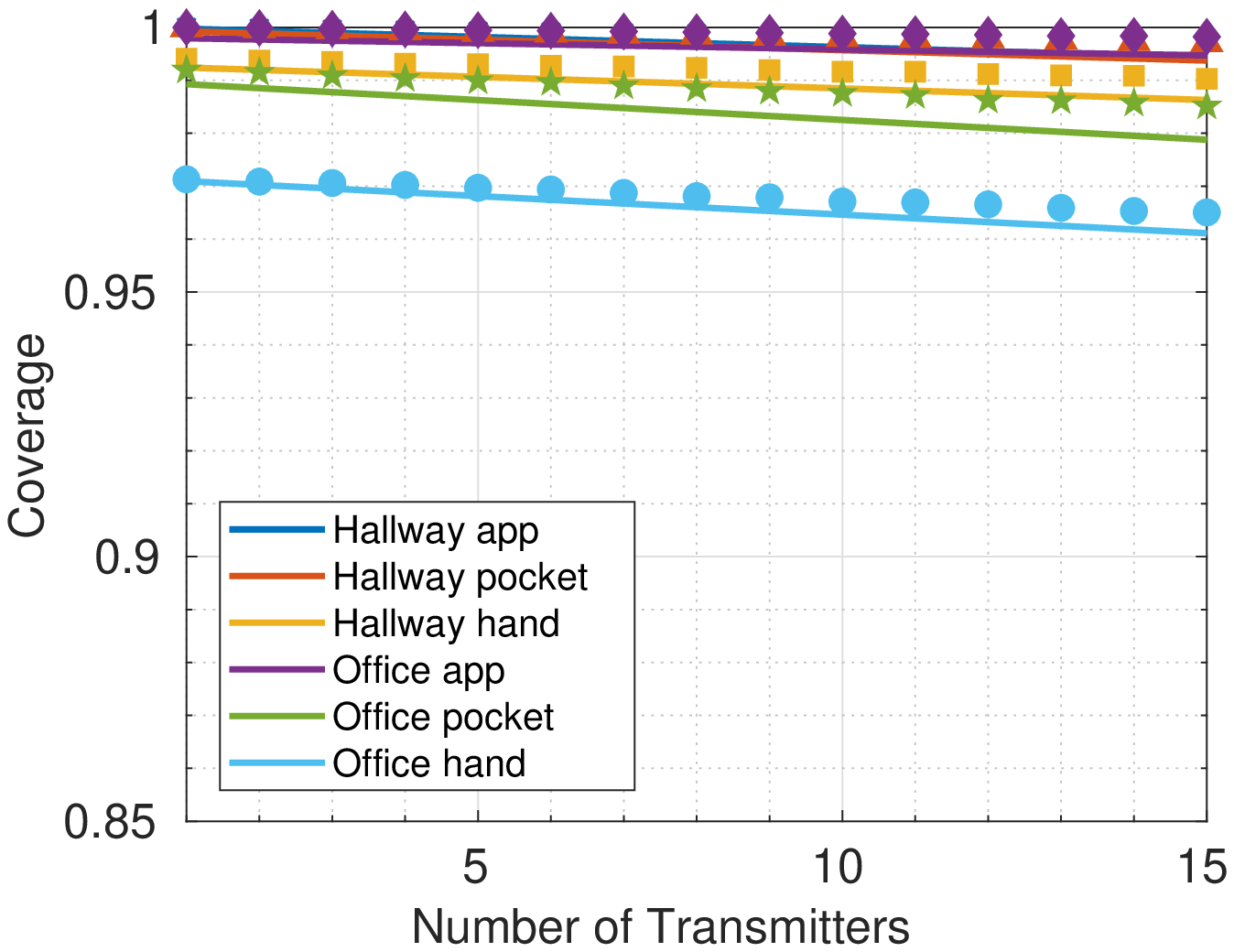}
}
\caption{Impact of the number of transmitters $n_{\tx}$. Where $n_{\tx}=1$ represents the scenario with a single serving transmitter and no other transmitters within a given area. Other network settings are: $\omega_{\tx}=\text{\unit[30]{$\degree$}}$, $r_0=\text{\unit[1]{m}}$, and $p_{\los}=0.5$.}
\label{fig:ntx}
\end{figure*}

First let us consider how the results we obtained compare to the performance requirements imposed on the \ac{3GPP} Indoor Hotspot scenario. Following the chairman of the \ac{3GPP} system architecture group \cite{norp20185g}, these include \unit[1]{Gbit/s} of \ac{EDR}, and \unit[15]{$\text{Tbps/km}^2$} of \ac{ATC}, both in the downlink direction. As we can see from \Fig{ntx}, whether or not we meet the performance targets is highly dependent on the usage scenario. Broadly speaking, scenarios where the user device is held further away from the user body achieve the \ac{ATC} target with 4 transmitters in a given area and maintain values higher than target \ac{EDR} for all of the considered network densities. The reason for this is low path attenuation at the reference distance. For that same reason Hallway pocket scenario meets target requirements. The other scenarios require double the number of transmitters to achieve the same performance, but, critically, do not offer target \ac{EDR}.

Looking at \Fig{ntx_atc} we see that the increasing number of transmitters leads to linear increase in \ac{ATC}, which, given the expression in \Eq{atc_def}, comes from the linear increases in the number of available infrastructure per unit area. Due to relatively narrow beams that we use, $\omega_{\tx}=30\degree$, this densification does not lead to increase in interference, which can be confirmed by observing almost flat \ac{EDR} curves in \Fig{ntx_edr} (which illustrates the performance of users that would suffer the most if any increase in interference occurred). The coverage plots in \Fig{ntx_cov} show the exact same story providing us with almost flat lines for each of the considered scenarios. In conclusion, we see that for the selected network setup our network operates in the noise-limited regime for all of the usage scenarios with the exception of the \textit{Hallway app} that is characterized by low reference distance path attenuation.

\begin{figure*}[tb!]
\centering
\subfigure[ATC\label{fig:beamwidth_atc}]{
 \includegraphics[width=0.31\textwidth]{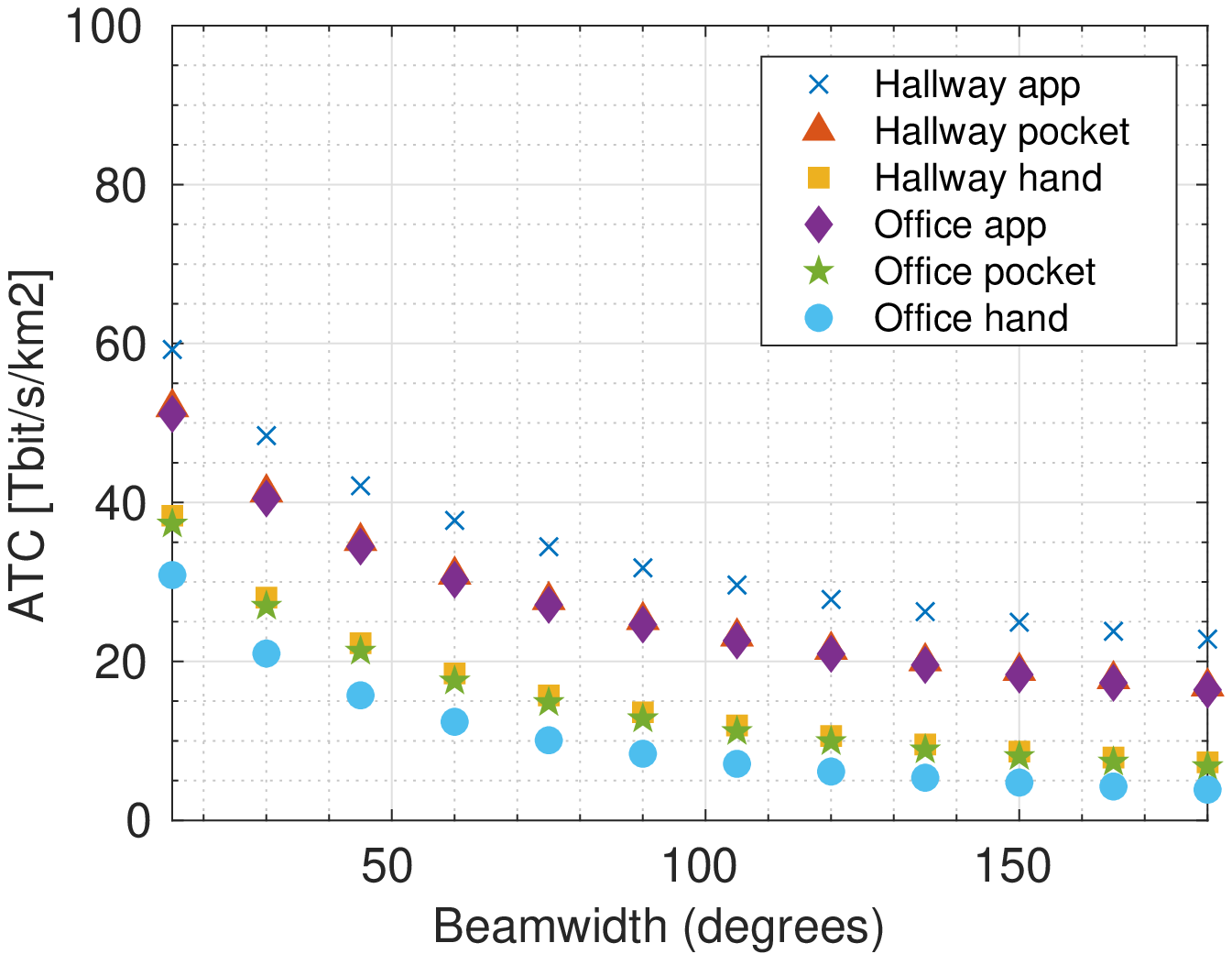}
}
\subfigure[EDR\label{fig:beamwidth_edr}]{
 \includegraphics[width=0.31\textwidth]{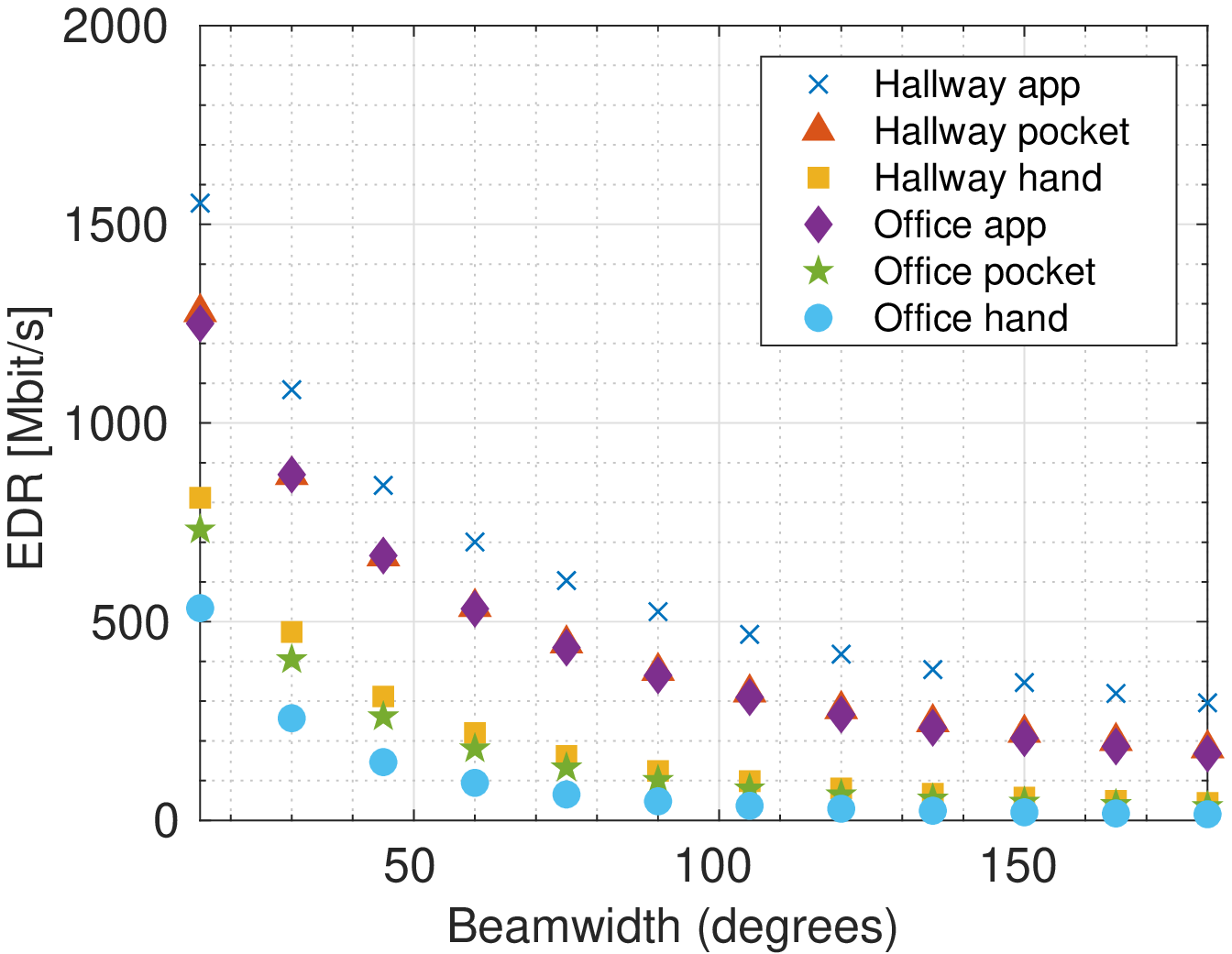}
}
\subfigure[Coverage\label{fig:beamwidth_cov}]{
 \includegraphics[width=0.31\textwidth]{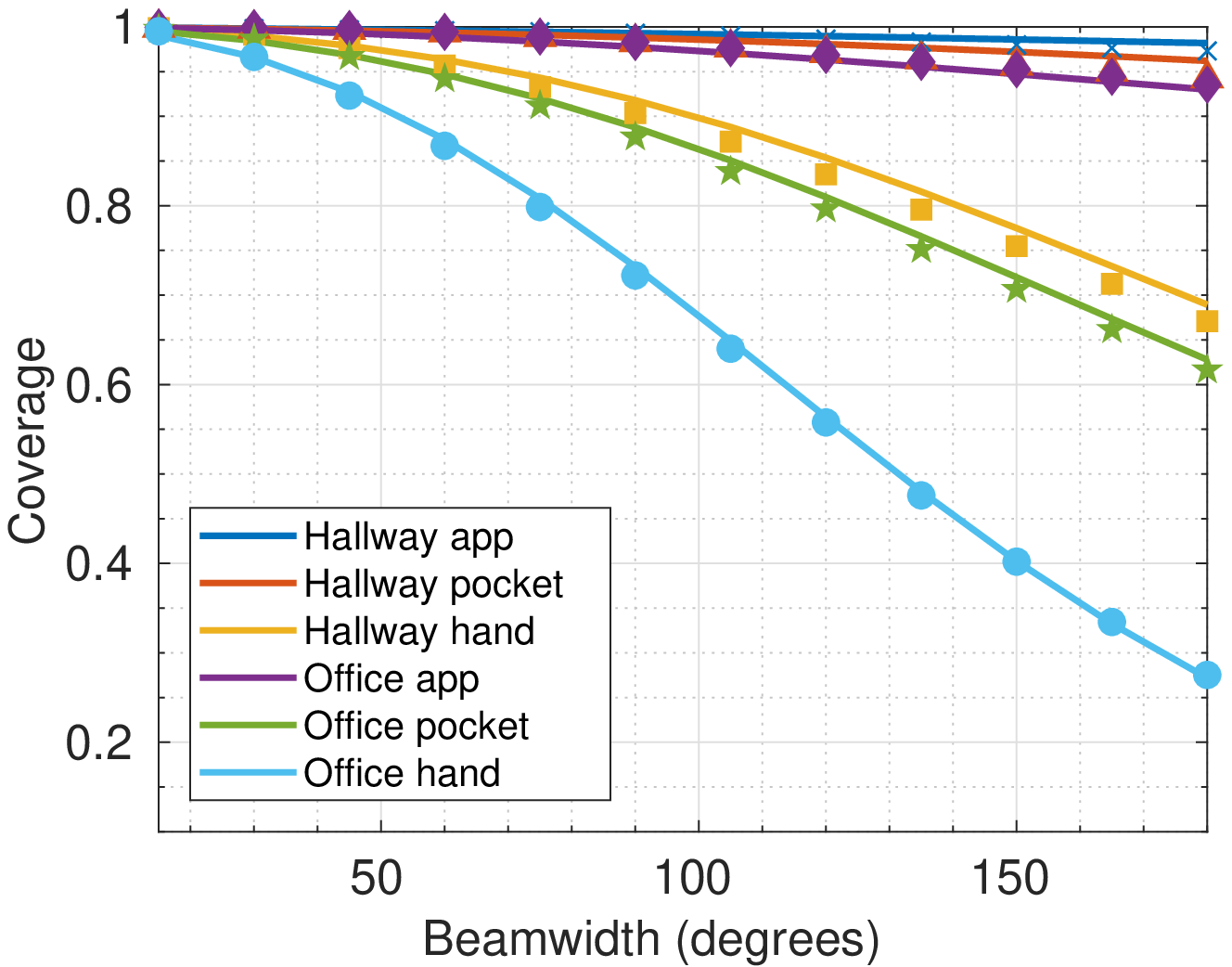}
}
\caption{Impact of the transmitter antenna main-lobe width $\omega_{\tx}$. The antenna gain scales with $\omega_{\tx}$ according to \Eq{mainlobe_gain}. Other network settings $n_{\tx}=12$, $r_0=\text{\unit[1]{m}}$, and $p_{\los}=0.5$.}
\label{fig:beamwidth}
\end{figure*}

Another way to bring the data rates up would require that we increase the antenna gains by using narrower beams, either on the transmitter (as shown in \Fig{beamwidth}), or the receiver side. Beamwidth has a much more critical impact on both the \ac{ATC} and \ac{EDR}. In \Fig{beamwidth_edr}, we can see that for some of the scenarios the \ac{EDR} drops to values below what, for example, an average \ac{LTE} user would experience. In order to ensure that users under all usage scenarios enjoy target \ac{EDR} would require that we use transmitter beamwidths far below our default configuration of \unit[30]{$\degree$}. Wider beams produce significant interference which greatly reduces the performance of the worst-off users for the three scenarios with high reference distance path attenuation. However, in the other three scenarios while the performance reduces it does so only to a lesser degree and \ac{ATC} stays above the target performance for all beamwidths considered. 

When it comes to coverage (see \Fig{ntx_cov} and \Fig{beamwidth_cov}) we should remark that---as one would expect---increasing the number of transmitters deteriorates coverage. In \Fig{ntx_cov}, coverage degrades linearly with the number of transmitters, albeit at a negligibly small rate. This is good news as it means that the resulting interference may be low enough to not warrant the need for interference coordination, at least as long as directionalities of transmissions are independent across space, an observation also made for large-scale, outdoor \ac{mmWave} networks \cite{GuptaAndrewsHeath_2015}. The choice of the beamwidth has a more critical impact on the coverage performance, in \Fig{beamwidth_cov}. If the coverage is to be kept at (or above) 90\% mark for all of the scenarios, it is necessary that the transmitter beamwidth stays at (or below) roughly \unit[60]{$\degree$}, as wider beams may produce unmitigated interference. Interestingly, when beams of \unit[50]{$\degree$} or more are used, we start seeing differences in coverage between the scenarios. Similarly to changes in \ac{ATC} and \ac{EDR} performance this can be explained by high discrepancy in the path attenuation at the reference distance between the usage scenarios.

\subsubsection{Network performance under design imperfections}

\begin{figure*}[tb!]
\centering
\subfigure[ATC\label{fig:serving_dist_atc}]{
 \includegraphics[width=0.31\textwidth]{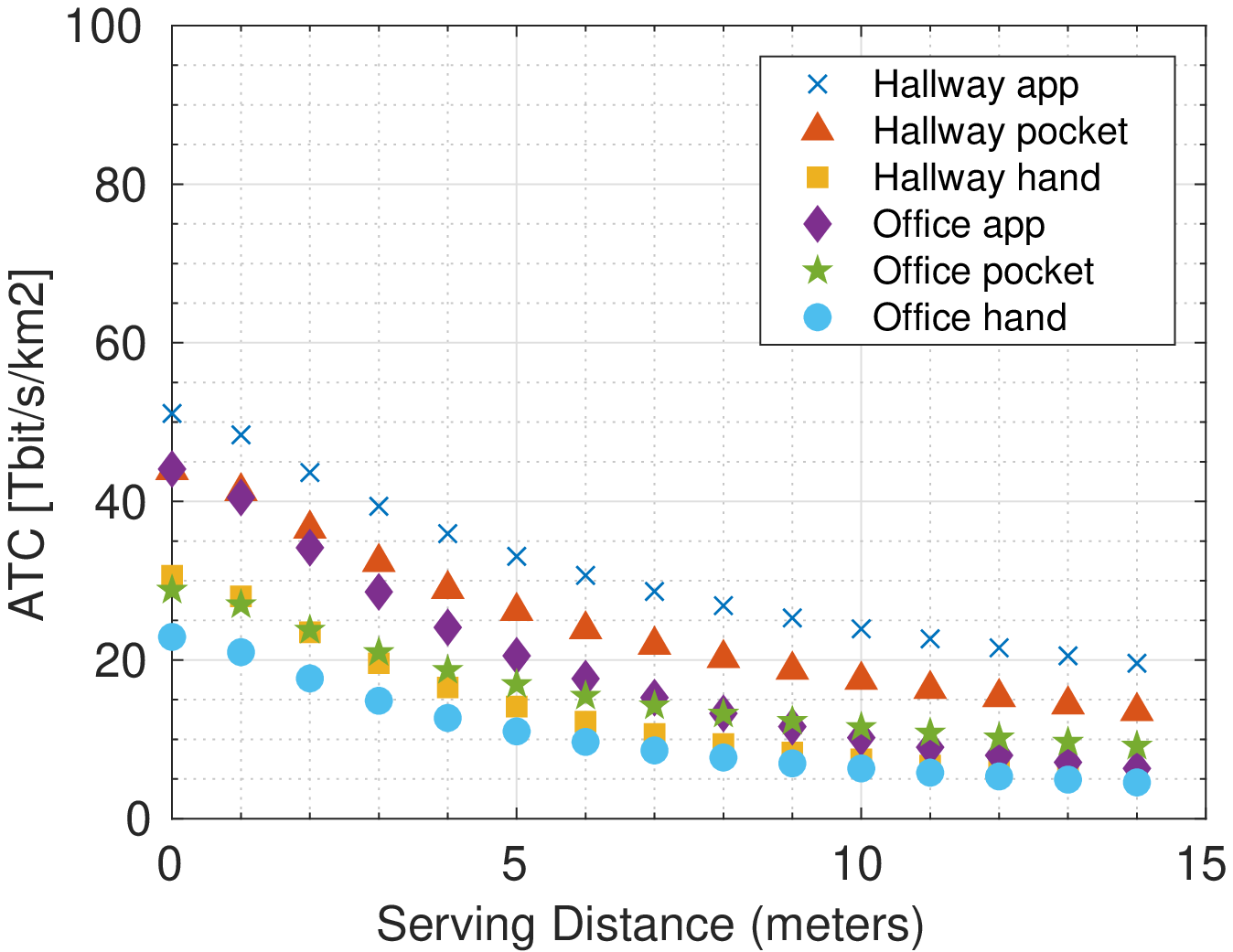}
}
\subfigure[EDR\label{fig:serving_dist_edr}]{
 \includegraphics[width=0.31\textwidth]{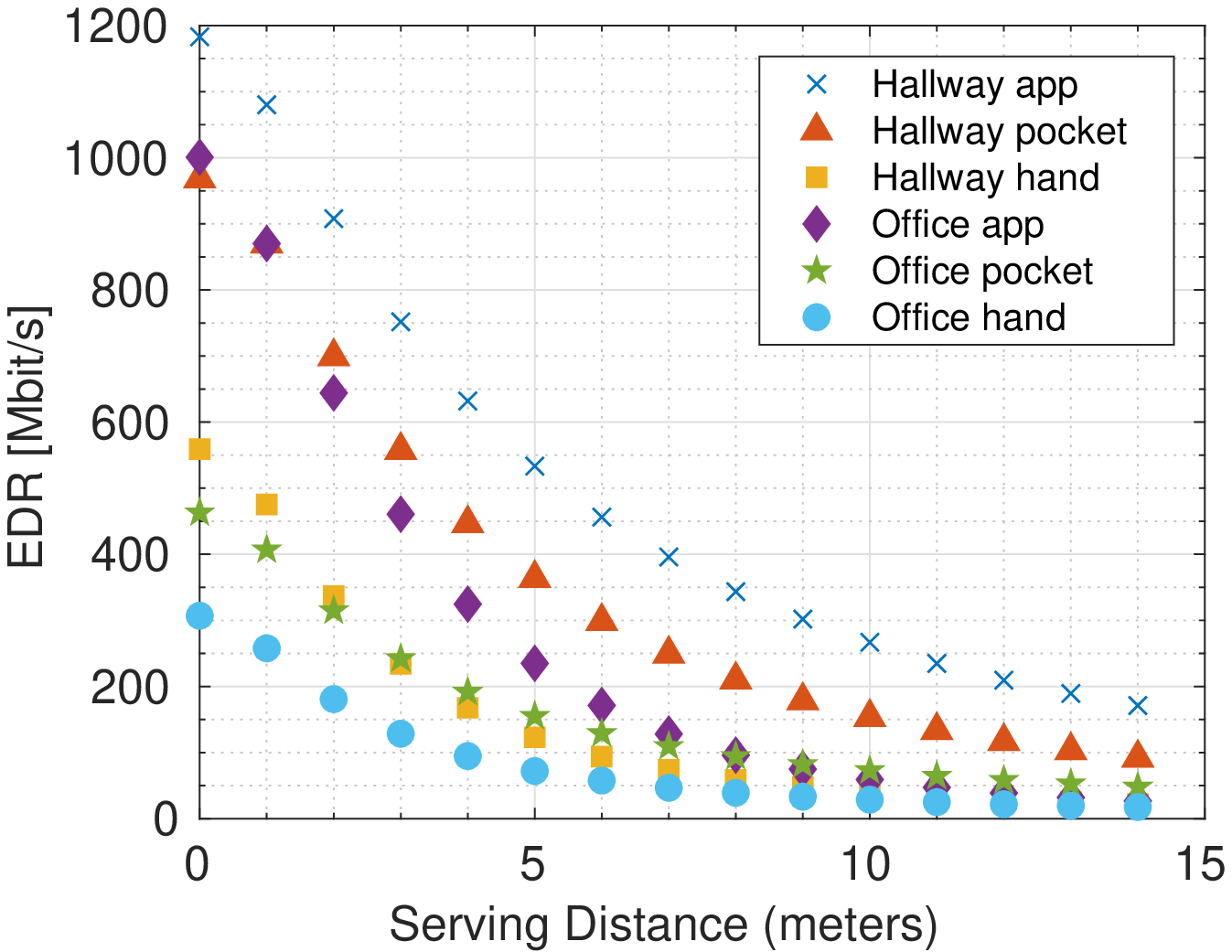}
}
\subfigure[Coverage\label{fig:serving_dist_cov}]{
 \includegraphics[width=0.31\textwidth]{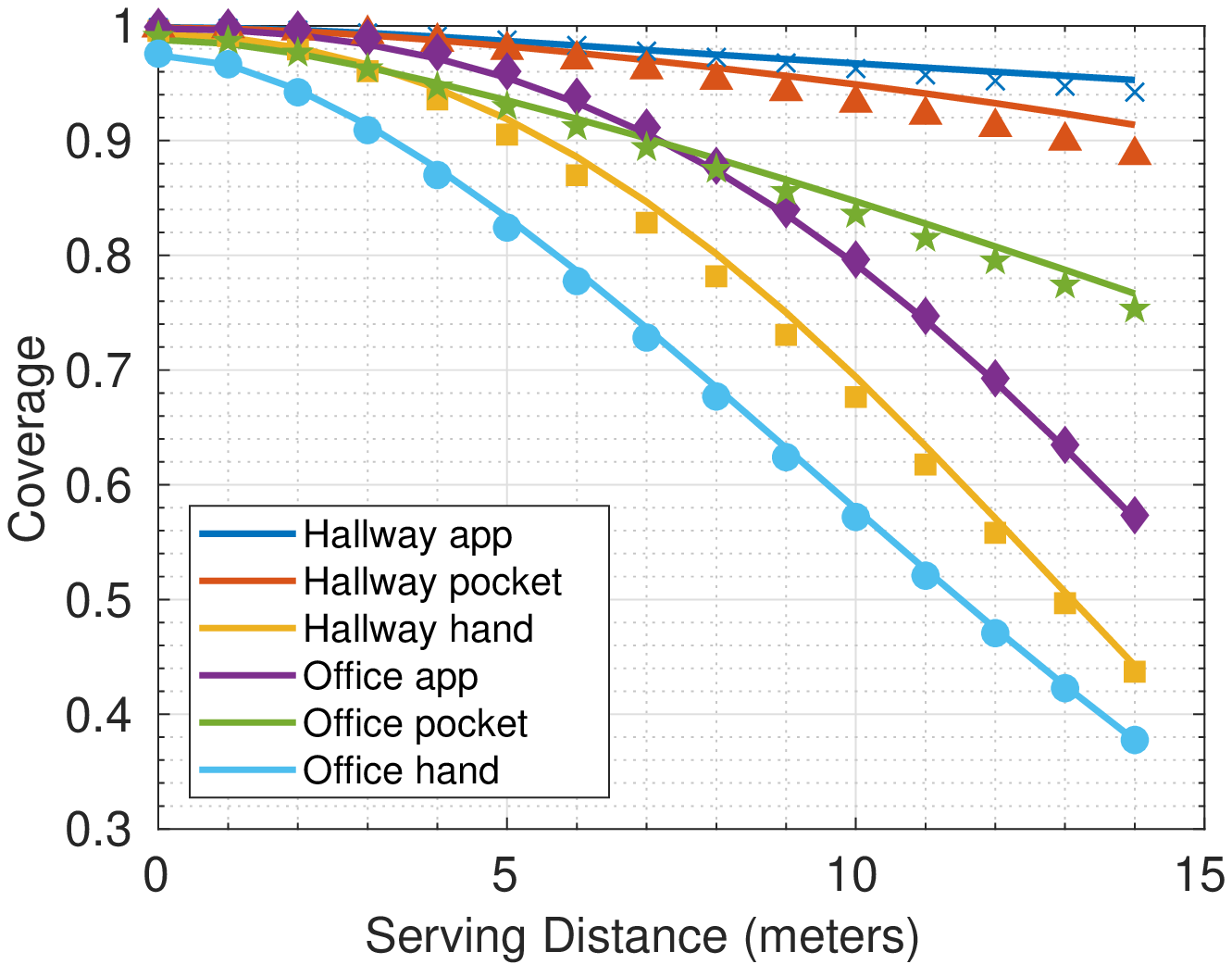}
}
\caption{Impact of the serving distance. Other network settings $n_{\tx}=11$, $\omega_{\tx}=\text{\unit[30]{$\degree$}}$, and $p_{\los}=.5$.}
\label{fig:serving_dist}
\end{figure*}

In \Fig{serving_dist} we consider the horizontal distance to the serving transmitter. In practical cases a particular cell will serve a particular user. This means the serving distance may not depend on the density of transmitters or receivers. In \Fig{serving_dist} we see that all our performance characteristics degrade with the serving distance. The serving distance is especially critical for the coverage, as in four of our usage cases the coverage goes below the 90\% mark at roughly \unit[5]{m} horizontal distance already. Interestingly this does not affect the \ac{ATC} performance which stays at relatively high values even at longer serving distances. Yet, the performance of worst-off users becomes highly volatile to change in their distance to the serving transmitter, and all of our scenarios fall below the \unit[1]{Gbit/s} target beyond the horizontal distance of \unit[2]{m}.

\begin{figure*}[tb!]
\centering
\subfigure[ATC\label{fig:los_atc}]{
 \includegraphics[width=0.31\textwidth]{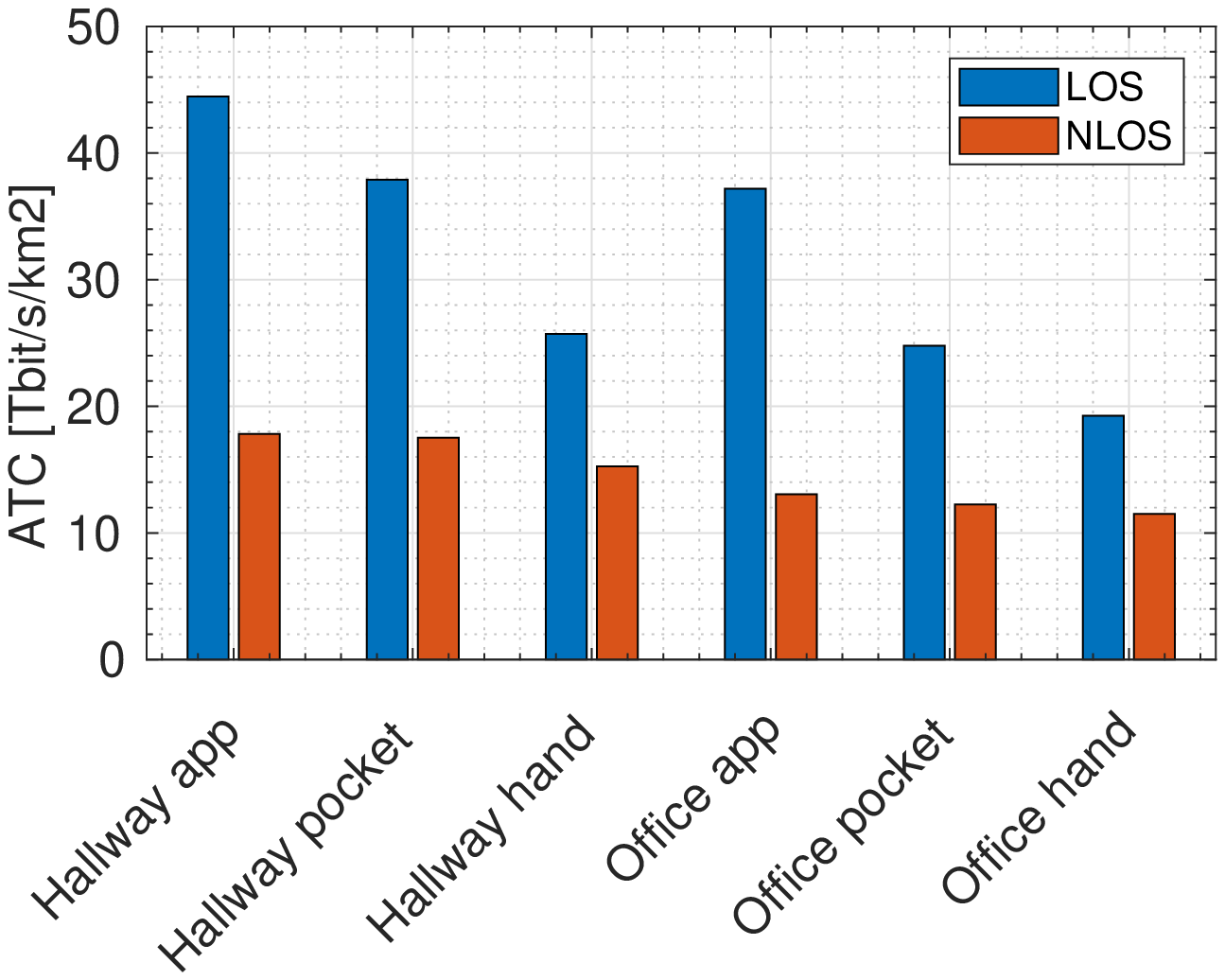}
}
\subfigure[EDR\label{fig:los_edr}]{
 \includegraphics[width=0.31\textwidth]{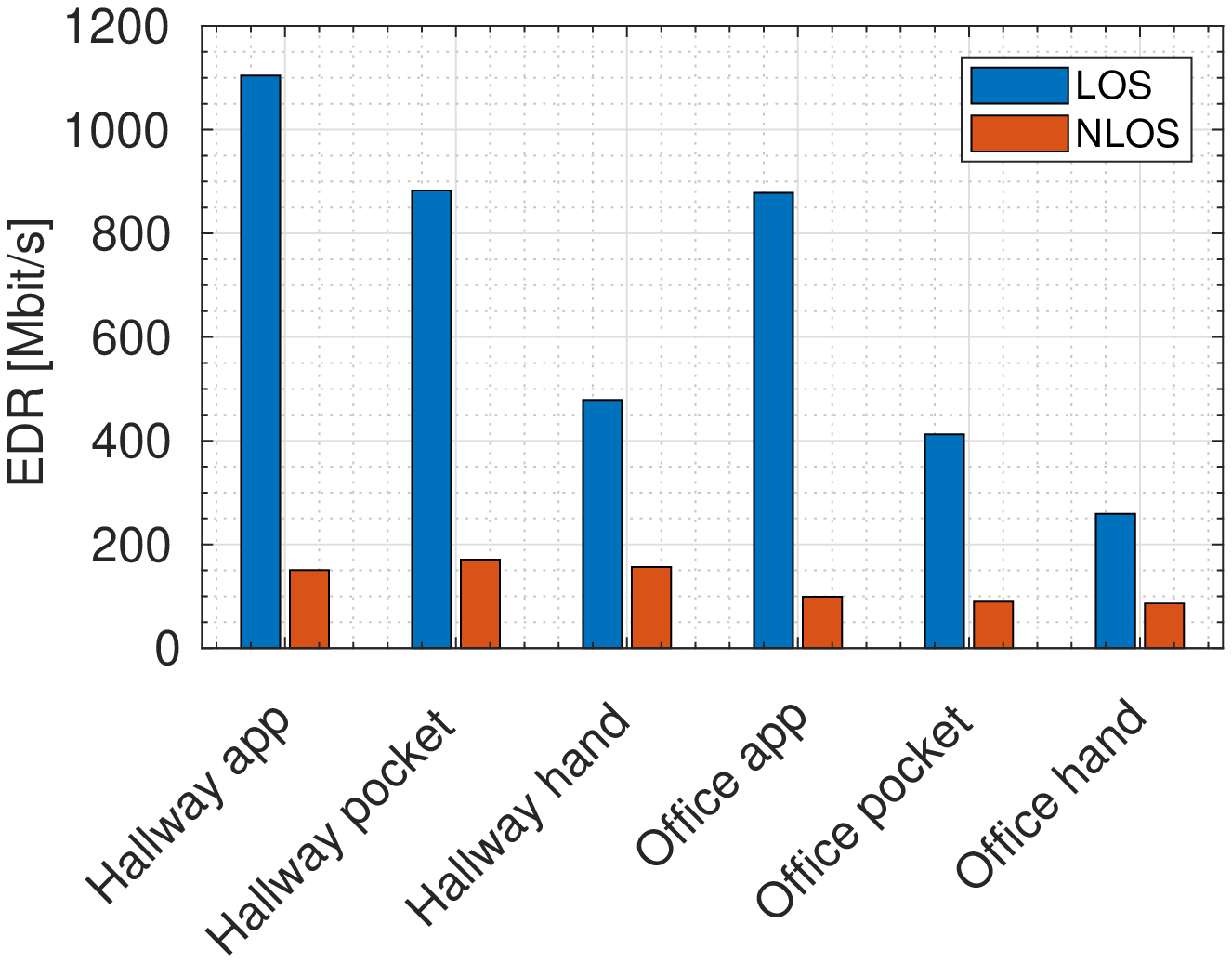}
}
\subfigure[Coverage\label{fig:los_cov}]{
 \includegraphics[width=0.31\textwidth]{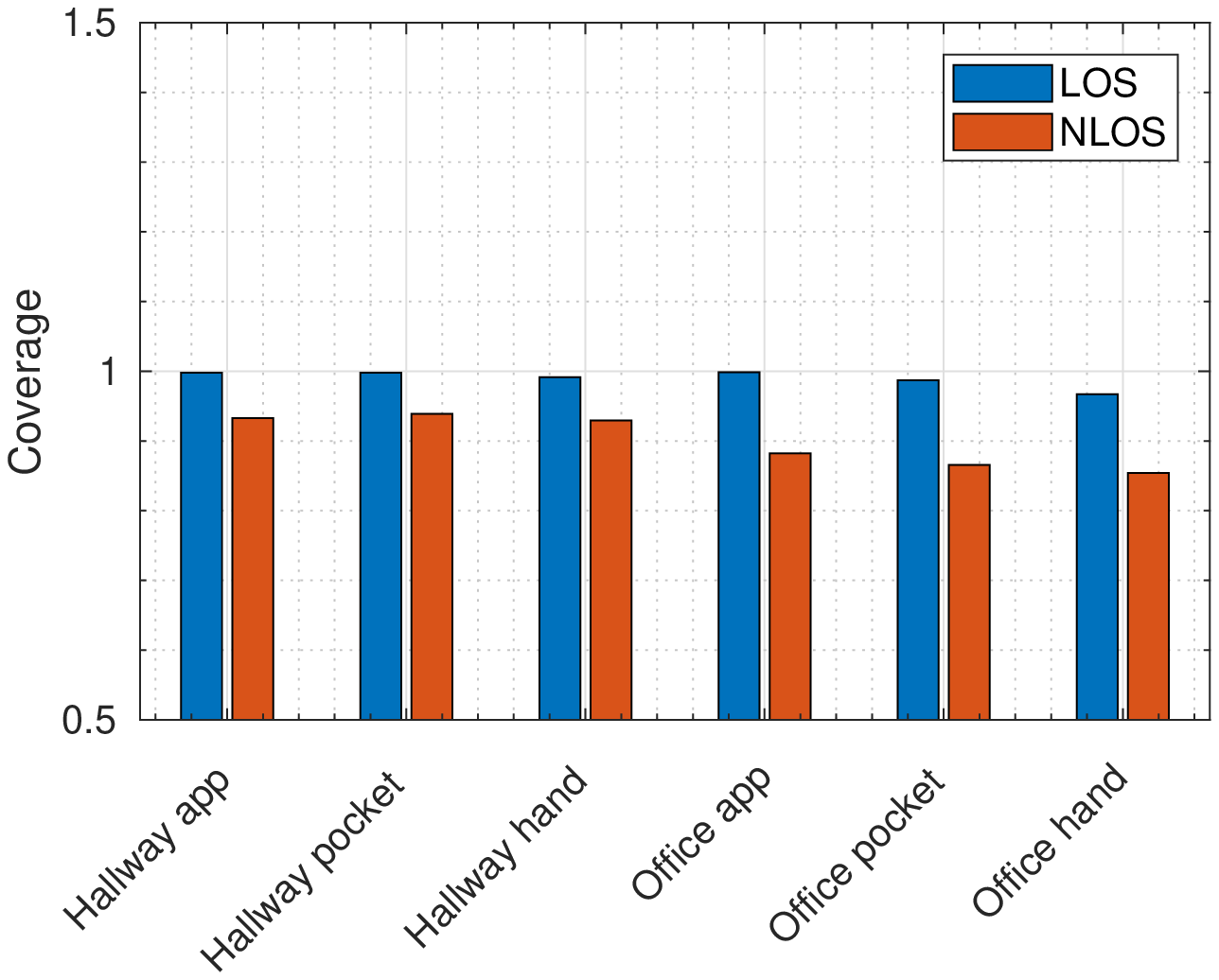}
}
\caption{Impact of the reference link blockage. Other network settings $n_{\tx}=11$, $\omega_{\tx}=\text{\unit[30]{$\degree$}}$, $r_0=\text{\unit[1]{m}}$, and $p_{\los}=.5$.}
\label{fig:los}
\end{figure*}

We are also interested in testing how the performance changes when the reference link is in blockage state caused by the user body. In \Fig{los} we see -- following intuition -- that \ac{LOS} blockage leads to performance degradation. While the coverage is affected only in a minor way, for each of the scenarios network performance, interpreted as \ac{ATC}, drops below the performance targets, in the cases where body blockage is high (Hallway app, Hallway pocket, and Office app) dropping by as much as 70\%. But more critically, the 5\textsuperscript{th} percentile user throughput becomes only a small fraction of the \ac{LOS} case. This is an important observation that motivates work into blockage mitigation strategies:  \cite{tatino2018maximum,petrov2018achieving,FiryagunaKibildaMarchetti_2019}. In our studies we also considered the impact of blockage probability. However, for the setup we used to evaluate our system, the disparity between \ac{LOS} and \ac{NLOS} channels on the interfering links was not strong enough to yield any significant differences in performance. We can conclude that while body blockage does indeed introduce significant attenuation to interfering links (see body blockage values in \Tab{para_estimate}), in a well-designed network with the carefully chosen beamwidths and a reasonable number of transmitters it should not affect network performance.

Since beam management is also a fundamental aspect to consider in \ac{mmWave} systems \cite{3GPP38802}, in our numerical evaluations we also considered the impact of beam misalignments. Given our system setup, we have observed that any misalignment between transmitter and receiver beams substantially throttles any communication links between the two, reducing coverage to almost zero. While this result motivates strong need for accurate beam alignment procedures, such as \cite{alkhateeb2017initial}, it is also based on a pessimistic antenna pattern model that incorporates sharp degradation in performance in case of any misalignments. With realistic antenna patterns, misalignments should lead to a more gradual performance degradation \cite{rebato2018study}, impact of which would require further studies.

\section{Deployment Analysis}
\label{sec:deployment}
\Sec{system_level} has addressed system-level modeling and performance evaluation. These tasks are critical to understanding the basic factors that will impact the performance of \ac{mmWave} deployments across a variety of scenarios. They help us to answer questions about the impact of factors such as base station density, antenna beamwidth, and serving distance on various measures of performance. Nevertheless, after examining these results, actual deployment planning for specific environments remains a challenge in \ac{mmWave} systems.

In a particular environment, deployment is constrained first by the set of possible \ac{AP} locations. Then, it is necessary for deployment to consider many of the same factors considered in the analysis in \Sec{system_level}. Namely, issues such as beamwidth and planned serving distance must be considered. In addition, we saw in the last section that whether a link is \ac{LOS} or \ac{NLOS} has a significant impact on system performance. For a particular deployment, the issue of \ac{LOS} vs. \ac{NLOS} will depend on the location of the \acp{AP}, user device orientation, as well as orientations of all human bodies in the considered space, including that of the device user. 

In this section, we discuss efficient schemes for \ac{AP} deployment and beam steering in \ac{mmWave} networks. In particular, given a set of possible \ac{AP} locations, we setup a stochastic optimization problem to determine the best set of \ac{AP} locations, as well as the best directions to aim the \acp{AP}' beams in order to meet coverage requirements while minimizing the deployment cost. We use a stochastic optimization problem to model the fact that the set of user locations at any given time is uncertain, but can be modeled as the realization of a point process. 
For the purposes of this paper, we step beyond the recent work on \ac{mmWave} \ac{AP} deployment by assuming that each \ac{AP} generates multiple, fixed beams; dynamic beam management is left for future work. 

There have been some recent works on \ac{AP} deployment in \ac{mmWave} networks, such as~\cite{Szyszkowicz16, Ghadikolaei15, Yuzhe16, soorki2017joint}. However, these works neither consider the beam steering problem nor account for the uncertainty in user locations. In~\cite{Szyszkowicz16}, the authors proposed an automated scheme for placing \ac{mmWave} \acp{AP} and gathering their line-of-sight coverage statistics, to help model small-cell \ac{mmWave} access networks. Considering the deafness and blockage problems in mmWave networks, in~\cite{Ghadikolaei15, Yuzhe16} the authors proposed distributed schemes for association and relaying that improve network throughput. In another \ac{AP} deployment scheme, \cite{soorki2017joint}, the authors assumed that \ac{AP}s always direct their beams in one fixed direction and considered a fixed set of \acp{UE} with static locations. In contrast, we assume fixed beam directions, but we assume that each \ac{AP} can generate multiple beams. 

Considering uncertainty in the availability of \ac{mmWave} links between \ac{AP} and \acp{UE}, combined with user location uncertainty, in this section we describe a \ac{CCSP}~\cite{kall1994stochastic} framework for joint \ac{AP} \underline{d}eployment and \underline{b}eam steering in m\underline{mWave} networks, called DBmWave. \ac{CCSP} has been recently used to model several resource allocation problems in uncertain networks~\cite{MJ_TW_13,chatterjee2018virtualization,Gomez_19}. DBmWave aims at minimizing the required number of \ac{mmWave} \acp{AP} to achieve a minimum \emph{network-wide} coverage probability of $\beta$, which represents the requested \ac{QoS} level. The \emph{network-wide} coverage probability constraint formulated in this paper is in contrast to the \emph{per-user} coverage probability constraint formulated in~\cite{soorki2017joint}. Instead of formulating a constraint for each user to ensure that individual users are covered with a minimum probability $\beta$, we formulate a single constraint for the entire \ac{mmWave} network that ensures that any arbitrarily selected user will be covered with this minimum probability $\beta$. Using various reformulation techniques, we equivalently reformulate our stochastic program as a \ac{BLP}. Finally, we numerically analyze the performance of DBmWave under various system settings.

Note that in addition to user location, because of significant human body shadowing in \ac{mmWave} networks, user orientation is also a significant source of uncertainty. We have addressed this in some past work \cite{soorki2017joint} and may integrate such considerations into this work in the future. 

\subsection{System Model}\label{sec:system_model}

\begin{figure}
\centering
\includegraphics[width=0.5\columnwidth]{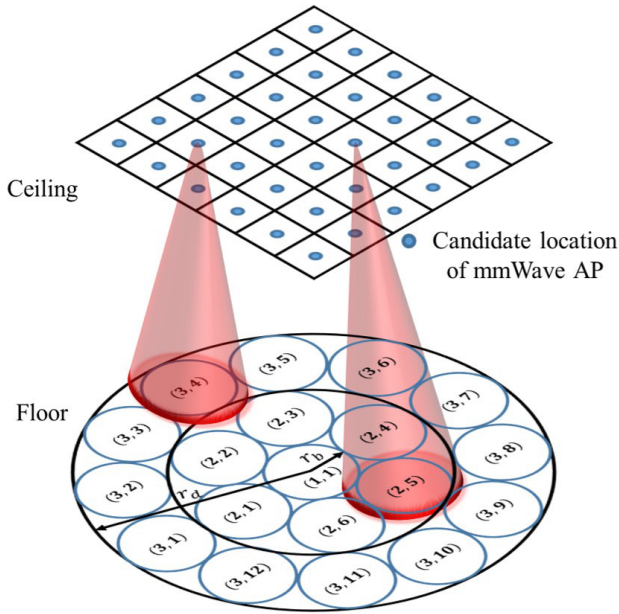}
\caption{Illustration of the system model considered in DBmWave.}
\label{fig:model}
\end{figure}

We consider a three-dimensional geographical area with a set $\mathcal{N} = \{1, 2, \ldots, N\}$ of candidate locations for deploying \ac{mmWave} \acp{AP} on the ceiling to cover the floor, as depicted in \Fig{model}. The floor is divided into $K = \frac{r_d}{2 \; r_b} + \frac{1}{2}$ annuli, the $i$th annulus consists of $M_i$ circles, where $M_i$ is given by: 
\begin{equation}
M_i = \Bigg\lfloor\frac{2 \pi}{2 \sin^{-1}\left(\frac{r_b}{r_d - 2 \; r_b (K - i) - r_b}\right)}\Bigg\rfloor = \Bigg\lfloor\frac{\pi}{\sin^{-1}\left(\frac{1}{2(i - 1)}\right)}\Bigg\rfloor.
\end{equation}

$r_d$ represents the radius of the geographical area and $r_b$ is the radius of each circular area, as depicted in \Fig{model}. We denote the set of circular areas by $\mathcal{K}$, where $|\mathcal{K}| = 1 + \sum_{i = 2}^{\frac{r_d}{2 \; r_b} + \frac{1}{2}} M_i$. The $k$th circular area in $\mathcal{K}$, denoted by $A_k$, is represented by a pair $(i_k, j_k)$, as illustrated in \Fig{model}.

\acp{UE} are distributed in the geographical area according to the distribution $f_Z(z)$. The link between a \ac{mmWave} \ac{AP} placed at location $n \in \mathcal{N}$ and the $k$th circular area, $k \in \mathcal{K}$, if one of the \ac{AP} beams is steered to cover $A_k$, is only available with probability $p_{nk}$. The maximum number of beams that a \ac{mmWave} \ac{AP} can have is denoted by $B$.

\subsection{Problem Statement}

\emph{Given $\mathcal{N}$, $\mathcal{K}$, $B$, $\beta$, $f_Z(z)$, and $p_{nk}, n \in \mathcal{N}, k \in \mathcal{K}$, we answer the following questions jointly while ensuring that an arbitrarily chosen user within the geographical area of interest will be covered with a probability $\geq \beta \in (0, 1)$.}
\begin{enumerate}
\item \emph{What is the minimum number of required \ac{mmWave} \acp{AP}?}
\item \emph{How can they be deployed optimally?}
\item \emph{How can their beams be steered optimally?}
\end{enumerate}

\subsection{Problem Formulation}

Let $y_{nk}, n \in \mathcal{N}, k \in \mathcal{K}$, be binary decision variables; $y_{nk}$ equals one if a \ac{mmWave} \ac{AP} is placed at location $n$ and one of its beams is steered to cover region $k$, and it equals zero otherwise. Let $\mathbb{P}_{\text{cov}}$ be the network-wide coverage probability, i.e., the probability that an arbitrarily selected user in the network will be covered. Then, the joint \ac{AP} deployment and beam steering problem can be formulated as: 
\begin{center}
\begin{tcolorbox}[title = Problem 1: Joint \ac{AP} Deployment and Beam Steering, width = 4in]
\begin{align}
& \underset{\left\{\substack{y_{nk}}\right\}}{\text{min}} \; \sum_{n \in \mathcal{N}} \; \mathds{1}_{\left\{\sum_{k \in \mathcal{K}} y_{nk} \geq 1\right\}} \label{eqn:objective1}\\
& \text{subject to:}  \nonumber \\
& \hspace{0.7in} \mathbb{P}_{\text{cov}} \geq \beta \\
& \hspace{0.7in} \sum_{k \in \mathcal{K}} y_{nk} \leq B, \forall n \in \mathcal{N} \\
& \hspace{0.7in} y_{nk} \in \{0, 1\}, \forall n \in \mathcal{N}, \forall k \in \mathcal{K} 
\end{align}
\end{tcolorbox}
\end{center}
\noindent where $\mathds{1}_{\{\cdot\}}$ is an indicator function; $\mathds{1}_{\{\cdot\}}$ equals one if $\{\cdot\}$ is satisfied and zero otherwise, and $\beta \in (0, 1)$.

\subsubsection{Coverage Probability Constraint}

As stated earlier, the coverage probability is defined as the probability that an arbitrarily selected user lies in a circular area that is covered by at least one active beam. Hence, the coverage probability when the arbitrarily selected user is located at $z$ can be defined as:
\begin{align}
\mathbb{P}_{\text{cov}}^{(z)} = \mathbb{E}\left[1 - \prod_{n\in\mathcal{N}}\left(1 - \delta_{nk^{(z)}}y_{nk}\right)\right],\label{eq:CPC}
\end{align}
where $k^{(z)}$ is the index of the circular area that contains location $z$. $\delta_{nk^{(z)}}$ equals one if there is no blockage between the \ac{AP} candidate location $n$ and the circular area $A_{k^{(z)}}$, and it equals zero otherwise. The expectation in~\Eq{CPC} is over blockages, which---similarly to \Sec{system_level}--- are assumed to be independent across links. Therefore,
\begin{align}
\mathbb{P}_{\text{cov}}^{(z)} = 1 - \prod_{n\in\mathcal{N}}\left(1-p_{nk^{(z)}}y_{nk}\right).
\end{align}

To compute the unconditioned coverage probability, we take the user distribution $f_Z(z)$ into consideration as follows:
\begin{align} 
\mathbb{P}_{\text{cov}} & = 1 - \sum_{k \in \mathcal{K}} \left(\int_{A_k} f_Z(z) \; {\rm d}z \prod_{n \in \mathcal{N}} \left(1 - p_{nk} \; y_{nk}\right)\right).
\end{align}

The integration $\int_{A_k} f_Z(z) \; {\rm d}z$ over each circular area is upper-bounded by the integration over the sector enclosed in the $i_k$-th annulus between the two tangents of $A_k$, see~\Eq{integral}. This upper bound is used in our analysis to enhance tractability. This enables us to use the probability distribution $f_{R_z}(r_z)$, where $R_z=\|z\|$. The term $\left(2 \sin^{-1}\left(\frac{1}{2(i_k - 1)}\right)\right)$ in~\Eq{integral} represents the angle of the sector enclosed in the $i_k$-th annulus between the two tangents of $A_k$. The term $a_{i_k}$ is added to ensure that $\sum_{k \in \mathcal{K}}\int_{A_k} f_Z(z) = 1$. In \Sec{Numerical_Analysis}, the following two user distributions are investigated: 
\begin{itemize}
\item \emph{Truncated Gaussian distribution}, where $f_{R_z}^{\rm gaus}(r_z)=\frac{r_z\exp\left(-\frac{r_z^2}{2\sigma^2}\right)}{\sigma^2\left(1-\exp\left(-\frac{r_d^2}{2\sigma^2}\right)\right)}$ and $\sigma^2$ represents the variance of the user distribution.
\item \emph{Uniform distribution}, where $f_{R_z}^{\rm unif}(r_z)=\frac{2r_z}{r_d^2}$.
\end{itemize}
\begin{figure*}
\begin{align}
\label{eq:integral}
\int_{A_k} f_Z(z) \; {\rm d}z & \leq \left(a_{i_k} + 2 \sin^{-1}\left(\frac{1}{2(i_k - 1)}\right)\right) \int_{r_d - 2 r_b (K - i_k + 1)}^{r_d - 2 r_b (K - i_k)}\frac{f_{R_z}(r_z)}{2\pi} \; {\rm d}r_z \nonumber \\
& = \left(a_{i_k} + 2 \sin^{-1}\left(\frac{1}{2(i_k - 1)}\right)\right) \int_{2 r_b (i_k - 1.5)}^{2 r_b (i_k - 0.5)} \frac{f_{R_z}(r_z)}{2\pi} \; {\rm d}r_z, \nonumber \\
 & \hspace{3in} \forall 2 \leq i_k \leq K.
\end{align}
\rule{\textwidth}{0.5pt}
\end{figure*}

\subsubsection{Equivalent \acl{BLP}}

First, note that the objective function of Problem 1 is non-linear. It can be represented in a linear form by introducing new binary decision variables, $x_n \triangleq \mathds{1}_{\left\{\sum_{k \in \mathcal{K}} y_{nk} \geq 1\right\}}, \forall n \in \mathcal{N}$, and reformulating the indicator function as follows~\cite{reformulation}:
\begin{itemize}
\item If $\sum_{k \in \mathcal{K}} y_{nk} \geq 1$ then $x_n = 1$ can be reformulated as:
\begin{equation}\label{eq:1}
\sum_{k \in \mathcal{K}} y_{nk} - \left(M + \epsilon\right) x_n \leq 1 - \epsilon,
\end{equation}
where $M$ is an upper bound of $\sum_{k \in \mathcal{K}} y_{nk} - 1$ and $\epsilon > 0$ is a small tolerance beyond which we regard the constraint as having been broken. Selecting $M$ and $\epsilon$ to be $B - 1$ and $1$, respectively,~\Eq{1} reduces to $\sum_{k \in \mathcal{K}} y_{nk} \leq B \; x_n$.
\item If $x_n = 1$ then $\sum_{k \in \mathcal{K}} y_{nk} \geq 1$ can be reformulated as\footnote{\scriptsize{Note that this condition is equivalent to $\sum_{k \in \mathcal{K}} y_{nk} = 0 \Longrightarrow x_n = 0$, which is already enforced by the objective function, since it aims at minimizing the number of \ac{mmWave} \acp{AP}. Hence,~\Eq{2} is redundant.}}:
\begin{equation}\label{eq:2}
\sum_{k \in \mathcal{K}} y_{nk} + m \; x_n \geq m + 1,
\end{equation}
where $m$ is a lower bound of $\sum_{k \in \mathcal{K}} y_{nk} - 1$. Selecting $m$ to be $-1$,~\Eq{2} reduces to $\sum_{k \in \mathcal{K}} y_{nk} \geq x_n$.
\end{itemize}
Therefore, 
\begin{equation}
x_n = \mathds{1}_{\left\{\sum_{k \in \mathcal{K}} y_{nk} \geq 1\right\}} \; \Longleftrightarrow \;  x_n \leq \sum_{k \in \mathcal{K}} y_{nk} \leq B \; x_n, \forall n \in \mathcal{N}. \nonumber 
\end{equation}

Second, the coverage probability expression, $\mathbb{P}_{\text{cov}}$, has the term $\mathbb{P} \triangleq \prod_{n \in \mathcal{N}} \left(1 - p_{nk} \; y_{nk}\right)$, which is nonlinear in the decision variables $y_{nk}, n \in \mathcal{N}, k \in \mathcal{K}$. Expanding $\mathbb{P}$, we can see that the nonlinear terms in $\mathbb{P}$ are in the form of products of binary decision variables. For example, if $N = 3$, $\mathbb{P}$ can be expressed as:
\begin{equation}
\mathbb{P} = 1 - \sum_{i=1}^3 p_{ik} \; y_{ik} + p_{1k} \; p_{2k} \; y_{1k} \; y_{2k} + p_{1k} \; p_{3k} \; y_{1k} \; y_{3k} + p_{2k} \; p_{3k} \; y_{2k} \; y_{3k} - \prod_{i=1}^3 p_{ik} \; y_{ik}.
\end{equation}

To linearize a product of binary decision variables, say $\prod_{i=1}^N y_{ik}$, we introduce a new auxiliary non-negative decision variable, say $y_k$, replace $\prod_{i=1}^N y_{ik}$ by $y_k$, and add the following constraints:
\begin{align}
& y_k \leq y_{ik}, \forall i \in \{1, 2, \ldots, N\} \nonumber \\
& y_k \geq \sum_{i=1}^N y_{ik} - \left(N - 1\right) \nonumber \\
& y_k \geq 0.
\end{align}    

After reformulating the indicator function and $\mathbb{P}$, as explained above, Problem 1 becomes a \ac{BLP}.

\subsection{Numerical Analysis}\label{sec:Numerical_Analysis}

\subsubsection{Setup}

Assuming an open indoor environment, $r_d$ and $r_b$ were selected to be 5.5 and \unit[0.5]{meters}, respectively. Based on these values, we calculated the number of circular areas, as explained in subsection~\ref{sec:system_model}, and found that $K = 92$. The maximum number of beams that a \ac{mmWave} \ac{AP} can have, $B$, is varied between $1$ and $4$. The \acp{AP} are assumed to be mounted on the ceiling, which is $10 \times 10$ m$^2$. The height of the ceiling is assumed to be \unit[3]{m}, similarly to our measurement setup in Sections~\ref{sec:channel} and~\ref{sec:system_level}. The \acp{AP} are deployed in a grid-based manner, similarly to the indoor hotspot scenario presented in \cite{ITU-RM2412_2017}, that is, each candidate location is with equal distance to each other, as shown in  \Fig{model}. Two different user distributions were considered: (i) Gaussian distribution with mean $u=0$ and variance $\sigma=10$ and (ii) uniform distribution. The probabilities of link availability were calculated assuming three channel effects: path loss, small-scale fading, and blockage (the event of a user having no \ac{LOS} with a certain \ac{AP}). We adopt the power-law path loss model, as presented in \Eq{pathloss}. For \ac{LOS} case, we set $P_0$ to $78.31$ dB and $n$ to $2.1$. For \ac{NLOS} case, we set $P_0$ to $95.39$ dB and $n$ to $3.5$. We set $d_0$ to $1$ m.

The small-scale fading is assumed to follow the $\kappa-\mu$ distribution given by~\Eq{kappamu}. For \ac{LOS} case, we set $\kappa$ to $2.80$, $\mu$ to $0.77$, and $\Omega$ to $1.16$. For \ac{NLOS} case, we set $\kappa$ to $0.92$, $\mu$ to $0.96$, and $\Omega$ to $1.23$. 

Similarly to \Sec{system_level}, blockages across the links between the \ac{mmWave} \acp{AP} and the coverage areas are modelled as independent and identically distributed (i.i.d.) Bernoulli random variables. $p_{LOS}$ denotes the probability that there is no blockage on a certain link between an \ac{AP} and a coverage area and $p_{NLOS} = 1 - p_{LOS}$ denotes the probability that there is a blockage on the link. In general, $p_{LOS}$ varies from one link to another; for our numerical results, $p_{LOS}$ is set to $0.5$ for all the links. Moreover,
given the results in \Fig{ntx} which showed negligibly small degradation of coverage in response to the increasing number of transmitters and similarity in system parameters, we can assume that frequency reuse factor is $\ll 1$, i.e., interference is negligible. The Gaussian noise spectral density is set to \unit[-174]{dBm/Hz}.

While linearizing Problem 1 above, we assumed that, for each user, there are only three \ac{AP} candidate locations that can cover it. These three \ac{AP} locations form the best (most available) \ac{AP}-user links (i.e., links with the highest $p_{n,k}$ values for a given $k$). We selected different values of $N$, the number of \ac{AP} candidate locations, to better characterize the behavior of the system. We evaluated our stochastic optimization framework in terms of the required number of \acp{AP} for different coverage probabilities $\beta$. The optimization problem was solved using \textsf{CPLEX}.

\subsubsection{Results}

\begin{figure*}[tb!]
\centering
\subfigure[\label{fig:gaus}]{
 \includegraphics[width=0.31\textwidth]{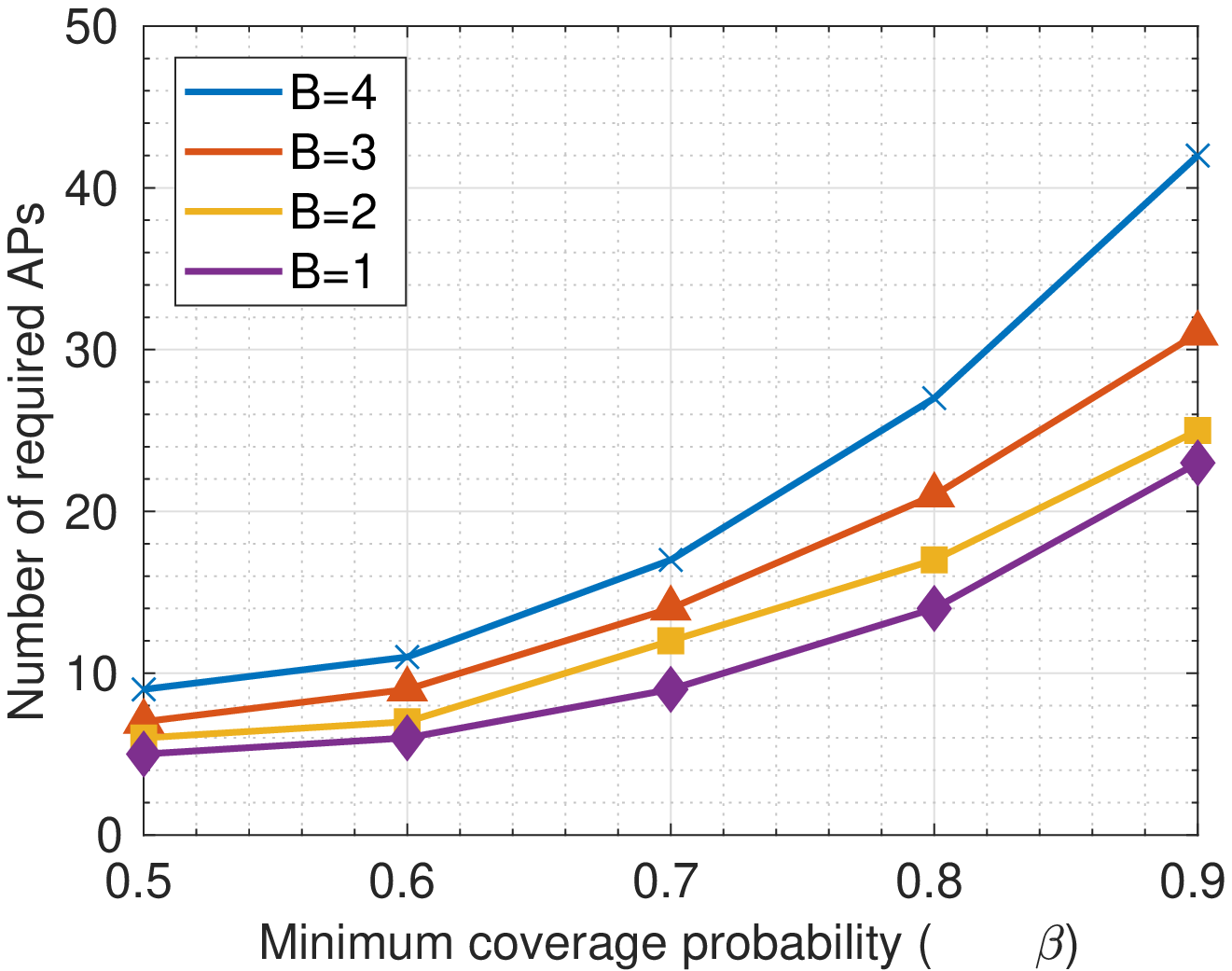}
}
\subfigure[\label{fig:uni}]{
 \includegraphics[width=0.31\textwidth]{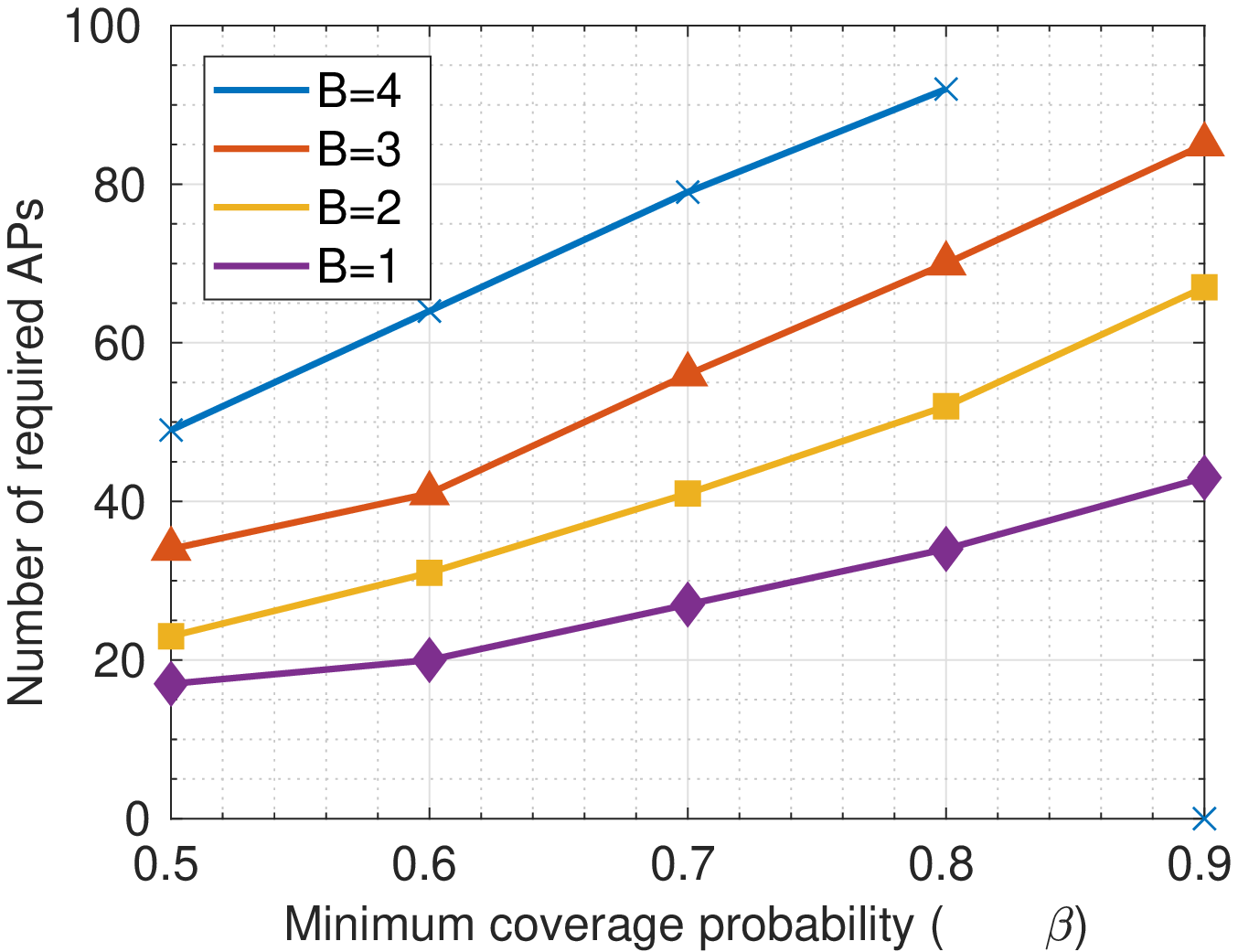}
}
\subfigure[\label{fig:nAP}]{
 \includegraphics[width=0.31\textwidth]{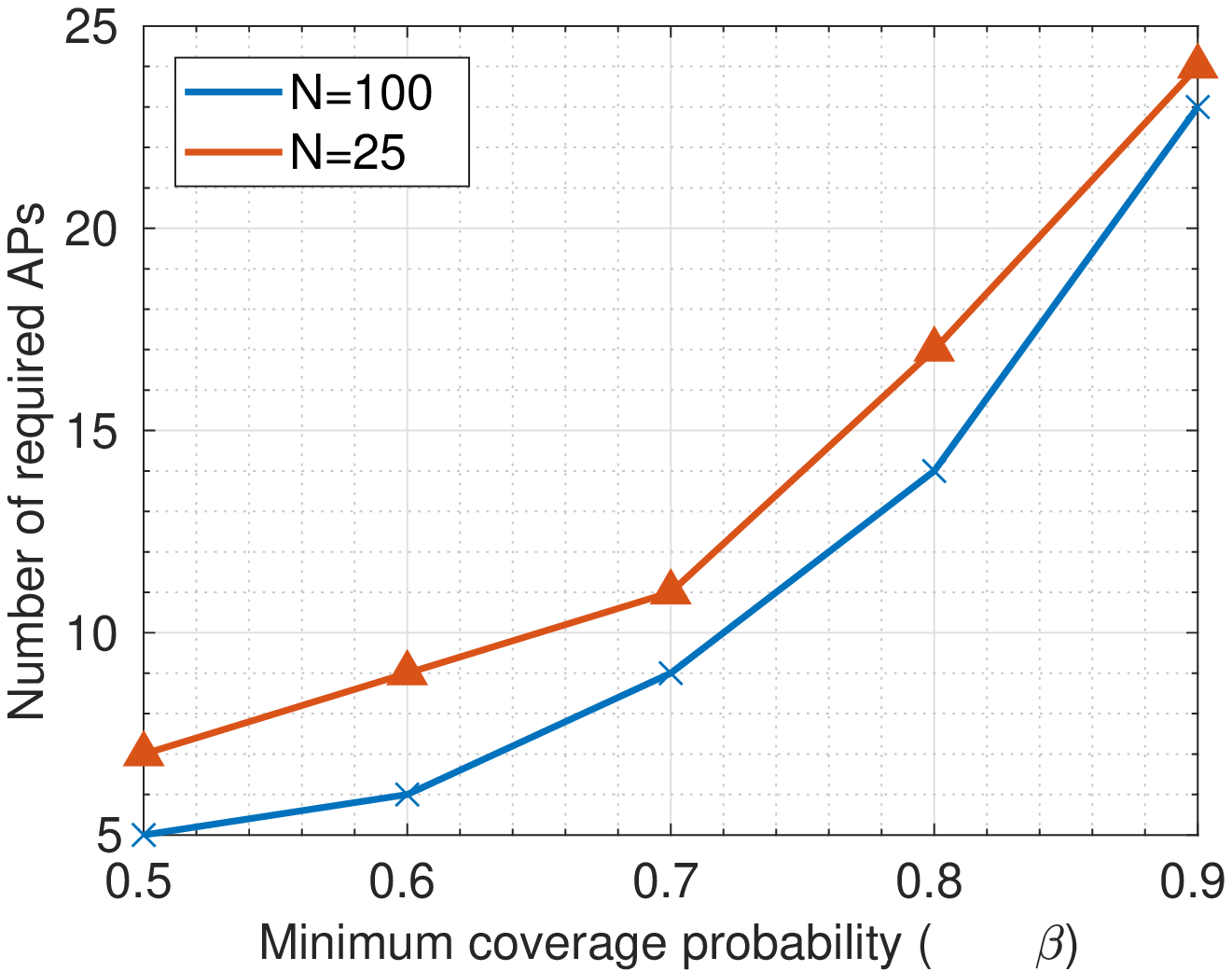}
}
\caption{Number of required \acp{AP} vs. minimum coverage probability for: a) Gaussian distributed users, with $N = 100$ beams, b) uniformly distributed users, with $N = 100$ beams, and c) Gaussian distributed users and different values of $N$, with $B = 4$.}
\label{fig:AP_deployment}
\end{figure*}

\Fig{gaus} shows the number of required \ac{AP}s as a function of the minimum required coverage probability ($\beta$). In this figure, the number of \ac{AP} candidate locations was chosen to be $N = 100$, the users were assumed to be distributed according to a Gaussian distribution. It can be seen that as $\beta$ increases, more \acp{AP} are needed to satisfy the coverage demand. Furthermore, increasing the number of beams at each \ac{AP} reduces the required number of \acp{AP}. This is expected, as having more beams at an \ac{AP} allows it to cover more users.

\Fig{uni} is similar to \Fig{gaus}, but assuming the users to be uniformly distributed. Both figures show similar trends. However, the number of \acp{AP} required to satisfy a certain coverage probability is higher when the users are uniformly distributed. In the case of Gaussian distribution, users are clustered in the geographical area (in contrast to the case of uniform distribution). This clustering results in reducing the number of required \acp{AP}.

Finally, \Fig{nAP} illustrates the effect of the number of \ac{AP} candidate locations on the number
of required \acp{AP} to meet a certain coverage probability. It can be seen that as $N$ increases the
number of required \acp{AP} decreases. This is due to the fact that increasing $N$ expands the feasibility
region of the allocation problem, opening the room for better solutions (i.e., with lower objective
function value).

\section{Discussion and Concluding Remarks}
\label{sec:future}

In the preceding sections we described findings that pertain to various aspects of indoor \ac{mmWave} network design. Crucially, we showed that indoor \acp{mmWave} deployments may achieve (or come close to achieving) performance targets of \unit[1]{Gbit/s} \ac{EDR}, and \unit[15]{$\text{Tbps/km}^2$} \ac{ATC} \cite{norp20185g}.
Yet, this performance can be volatile to device usage scenarios. In this section, we bring together the results to discuss these trade-offs, and look more closely at any outstanding challenges.

\subsection{Take Away Lessons}

In the system-level analysis, we saw that our reference user achieved almost 100\% coverage, even under the presence of body blockages, in \Fig{los_cov}. From the deployment analysis in \Fig{gaus} and \Fig{uni}, we saw that this does not need to be the case if multiple users are considered. Covering each and every user above a certain reliability threshold may require deploying an excessively large number of access points, especially if we expect high reliability. These additional transmitters will increase the network capacity, see \Fig{ntx_atc}, without deteriorating the rate for 5\textsuperscript{th} percentile users, see \Fig{ntx_edr}. However, depending on the usage scenario these 5\textsuperscript{th} percentile users may operate substantially below the required performance. In \Sec{deployment}, we showed that this can be avoided by coordinating deployment locations and beam steering according to the patterns of user locations (users that cluster in space) and user orientations (users that face single direction). This can be observed by comparing results in \Fig{gaus} and \Fig{uni}. 

Another way to bring up the 5\textsuperscript{th} percentile user performance is to increase the antenna gain by narrowing the transmission beamwidth, as we showed in \Fig{beamwidth_edr}. This will come at a cost of using more directional antennas or antenna arrays with larger number of components, and a potential reduction in the performance of beam tracking and alignment mechanisms, which require wider beams to operate (as reported in, e.g., \cite{alkhateeb2017initial}). This trade-off in beamwidth design can be avoided by planning for a deployment with less stringent coverage reliability requirements, which, as we could observe in \Fig{gaus}, greatly reduces the number of necessary transmitters, leading to reduced interference and enabling us the usage of wider beams.

\subsection{Future Challenges}

\subsubsection{Exploiting Differences in Channel Characteristics Across Multiple Frequency Bands}
One way to improve link availability in \ac{mmWave} networks is to make use of dual connectivity at microwave and \ac{mmWave} operating frequencies  \cite{3GPP37340}. To fully exploit the extra degree of freedom offered through dual-band operation, channel characteristics of both connections should be adequately de-correlated (i.e., if the \ac{mmWave} link goes into outage due to shadowing or a deep fade, then it would be anticipated, that under normal circumstances, the microwave link should act to provide a more reliable fall-back service). At present, we have a fairly good overall understanding of the quantitative differences in propagation within the microwave and \acp{mmWave} regions of the radio spectrum \cite{shafi2018microwave}. Nonetheless, important channel metrics such as the correlation in shadowing and multipath fading across frequency bands (introduced by human body blockages) is still largely uncharted territory, e.g., \cite{3GPP2}. To ensure that the instantaneous channel characteristics are suitably disparate for the envisaged multi-band operation, extensive channel measurement campaigns, as well as follow-up system-level analyses will be necessary.

\subsubsection{Coexistence between \ac{5G}-\ac{NR} and WiFi}
One of the key issues to address when operating in \ac{mmWave} unlicensed bands is coexistence with other wireless systems. For example, in the \unit[60]{GHz} band where the \ac{3GPP} is planning to deploy \ac{NR-U}, mechanisms will have to be developed to enable coexistence with wireless local area networks like the IEEE Wireless Gigabit 802.11ad/ay \cite{6979964_80211ad_magazine,8088544_80211ay_magazine}. As a matter of fact, the \ac{ETSI} has already published a list of conformance requirements, which include limits on the maximum emitted power or channel sensing mechanisms, necessary to ensure fair coexistence between the systems operating in unlicensed \ac{mmWave} bands \cite{ETSI_EN_302_567}. Yet, more work is needed to understand the impact of these coexistence rules on the network performance and, ultimately, network design. 

\subsubsection{Network Densification Limits}
Finally, in \Fig{ntx_atc}, we could see that the network capacity of our system increases with the number of transmitters. An interesting question is to ask about the asymptotic case, and the limits of network densification. In \cite{lopez2018brief}, it was observed that network densification is limited by the channel characteristics, and antenna heights. However, this result applies to networks operating over large areas and network settings that correspond to outdoor deployments, e.g., variable deployment heights of tens of meters above ground level. One would expect that for ceiling-mounted \ac{mmWave} networks, with highly directional beams, we can reach high deployment densities, without making the system interference-limited, and thus achieve high network capacities. Yet, this would make beam alignment and tracking more challenging, thus potentially limiting the achievable network capacity. Analysis of this network density asymptotic regime would likely require a new modelling framework.

\appendices

\section{Proof of \Lemma{recv_pwr_statistics}}
\label{app:kappamu_statistics}
The power received at the reference user (as provided in \Eq{rx_power}), conditioned on the blockage state $t$ and antenna gain $g_i$, is a product of the fading random variable and a constant. This allows us to express the \ac{PDF} of $S_{i,t}$ using \Eq{app1_kappa_mu} 
\begin{equation}
f_{S_{i,t}}(s) = \frac{\theta_1^{(\mu+1)/2}}{\vartheta_{i,t}^{(\mu+1)/2}\theta_2^{(\mu-1)/2}}s^{(\mu-1)/2}\exp\left(-\frac{\theta_1 }{\vartheta_{i,t}}s-\theta_{2,t}\right)I_{\mu_t-1}\left(2\sqrt{\frac{\theta_{1,t}\theta_{2,t} s}{\vartheta_{i,t}}}\right),
	\label{eq:app1_001}
\end{equation}
where $\vartheta_{i,t} = g_{i}l_{t}\left(r_i\right)$, and $I_{\mu_t-1}$ is the modified Bessel function of the first kind and order $\mu_t-1$.
Now, using the series representation of the modified Bessel function we get that
\begin{equation}
f_{S_{i,t}}(s) = \frac{\theta_{1,t}^{\mu_t}\exp(-\theta_{2,t})}{\vartheta_{i,t}^{\mu_t} }\sum_{l=0}^{\infty}\frac{\theta_{1,t}^l\theta_{2,t}^{l}}{ \vartheta_{i,t}^{l} \Gamma(\mu_t+l)l!}
s^{\mu_t-1+l}\exp\left(-\frac{\theta_{1,t}}{\vartheta_{i,t}} s\right).
\label{eq:app1_002}
\end{equation}

Then the \ac{CDF} of $S_{i,t}$ can be expressed as 
\begin{equation}
	\begin{split}
	\Pb\left(S_{i,t} > x \right) &= \int_{x}^{\infty} f_{S_{i,t}}(s) \dd{s} \\
    &\overset{(a)}{=} \frac{\theta_{1,t}^{\mu_t}\exp(-\theta_{2,t})}{\vartheta_{i,t}^{\mu_t} }\sum_{l=0}^{\infty}\frac{\theta_{1,t}^l\theta_{2,t}^{l}}{ \vartheta_{i,t}^{l} \Gamma(\mu_t+l)l!}\int_{x}^{\infty}
s^{\mu_t-1+l}\exp\left(-\frac{\theta_{1,t} }{\vartheta_{i,t}} s\right) \dd{s} \\ 
    &= \exp(-\theta_{2,t})\sum_{l=0}^{\infty}\frac{\theta_{2,t}^{l}}{ l! }\frac{\Gamma(l+\mu_t, \frac{\theta_{1,t} }{\vartheta_{i,t}}x)}{\Gamma(l+\mu_t)}\\
	&\overset{(b)}{=} \exp(-\theta_{2,t})\sum_{l=0}^{\infty}\frac{\theta_{2,t}^{l}}{l!} \sum_{n=0}^{l+\mu_t-1} \frac{1}{n!}\left( \frac{\theta_{1,t} x}{\vartheta_{i,t}} \right)^n \exp\left( -\frac{\theta_{1,t} }{\vartheta_{i,t}}x\right),
    \label{eq:app1_003}
	\end{split}
\end{equation}
where for (a) we use the expression in \Eq{app1_002}, and (b) holds only for the special case of $\mu_t$ being a positive integer.

\section{Proof of \Lemma{ccdf_bpp_network}}
\label{app:sinr_deterministic_network}
In the following, given the distance to the serving transmitter $r_0$, serving link gain $g_0$, and the serving channel being in state $t$, and denoting the longer-term average power received from the serving transmitter as $\vartheta_{0,t} = g_{0}l_{t}\left(r_0\right)$, we derive the conditional \ac{CCDF} of the \ac{SINR} as experienced by the reference user, i.e., $\Pb_{\text{cov}}=F^c_{\sinr|R_0,G_0,T}\left(\zeta|r_0,g_0,t\right)$. For a given threshold $\zeta$, this conditional \ac{CCDF} can be defined as 
\begin{equation}
\begin{split}
\Pb_{\text{cov}} &= \Pb\left(S > \zeta \cdot \left( I + \tau^{-1} \right)\right)\\
 &= \exp(-\theta_{2,t})\sum_{l=0}^{\infty}\frac{\theta_{2,t}^{l}}{l!} \sum_{n=0}^{l+\mu_t-1} \frac{\Eb[(\frac{\zeta\theta_{1,t}}{\vartheta_{0,t}} \left( I + \tau^{-1} \right))^n \exp\left( -\frac{\zeta\theta_{1,t}}{\vartheta_{0,t}} \left( I + \tau^{-1} \right)\right)]}{n!}\\
    &\overset{(a)}{=} \exp\left(-\theta_{2,t}-\frac{\zeta\theta_{1,t}}{\vartheta_{0,t}\tau}\right)\sum_{l=0}^{\infty}\frac{\theta_{2,t}^{l}}{l!}\sum_{n=0}^{l+\mu_t-1} \frac{1 }{n!} \left(\frac{\zeta\theta_{1,t}}{\vartheta_{0,t}}\right)^n  \sum_{k=0}^n {n\choose k} \tau^{-n+k} \Eb[I^k \exp\left( -\frac{\zeta\theta_{1,t}}{\vartheta_{0,t}}I \right)] \\
    &= \exp\left(-\theta_{2,t}-\frac{\zeta\theta_{1,t}}{\vartheta_{0,t}\tau}\right)\sum_{l=0}^{\infty}\frac{\theta_{2,t}^{l}}{l!}\sum_{n=0}^{l+\mu_t-1} \frac{1 }{n!} \left(\frac{\zeta\theta_{1,t}}{\vartheta_{0,t}\tau}\right)^n  \sum_{k=0}^n {n\choose k} \tau^{k}\\ 
    &\Eb\left[\left(\sum^{n_{\tx}}_{i=1}G_iH_{i,V}l_V(R_i)\right)^k \exp\left( -\frac{\zeta\theta_{1,t}}{\vartheta_{0,t}}\left(\sum^{n_{\tx}}_{i=1}G_iH_{i,V}l_V(R_i)\right) \right)\right] \\
    &\overset{(b)}{=} \exp\left(-\theta_{2,t}-\frac{\zeta\theta_{1,t}}{\vartheta_{0,t}\tau}\right)\sum_{l=0}^{\infty}\frac{\theta_{2,t}^{l}}{l!}\sum_{n=0}^{l+\mu_t-1} \frac{1 }{n!} \left(\frac{\zeta\theta_{1,t}}{\vartheta_{0,t}\tau}\right)^n  \sum_{k=0}^n {n\choose k} \tau^{k}\\ 
    &\sum_{k_1+k_2+\ldots+k_{n_{\tx}}=k}{k\choose k_1,k_2,\ldots,k_{n_{\tx}}}\prod_{1\leq i\leq n_{\tx}}\Eb\left[\left(G_iH_{i,V}l_V(R_i)\right)^{k_i} \exp\left( -\frac{\zeta\theta_{1,t}}{\vartheta_{0,t}}G_iH_{i,V}l_V(R_i) \right)\right]\\
	&\overset{(c)}{=} \exp\left(-\theta_{2,t}-\frac{\zeta\theta_{1,t}}{\vartheta_{0,t}\tau}\right)\sum_{l=0}^{\infty}\frac{\theta_{2,t}^{l}}{l!}\sum_{n=0}^{l+\mu_t-1} \frac{1 }{n!} \left(\frac{\zeta\theta_{1,t}}{\vartheta_{0,t}\tau}\right)^n  \sum_{k=0}^n {n\choose k} \tau^{k}\sum_{k_1+k_2+\ldots+k_{n_{\tx}}=k}{k\choose k_1,k_2,\ldots,k_{n_{\tx}}} \\
    &\prod_{1\leq i\leq n_{\tx}}\underbrace{\Eb_{V,G_i,R_i}\left[ \frac{(\mu_V)_{k_i} \theta_{1,V}^{\mu_V}\exp(-\theta_{2,V})\left(G_i l_V(R_i)\right)^{k_i}}{\left(\frac{\zeta\theta_{1,t}}{\vartheta_{0,t}}G_i l_V(R_i) + \theta_{1,V} \right)^{k_i+\mu_V}} \confhypergeom{k_i+\mu_V}{\mu_V}{\frac{\theta_{1,V}\theta_{2,V}}{\frac{\zeta\theta_{1,t}}{\vartheta_{0,t}}G_i l_V(R_i) + \theta_{1,V}}}\right]}_{\mathbb{I}_1},
	\end{split}
	\label{eq:sysdet_001}
\end{equation}
where (a) comes from the binomial expansion, in (b) we apply the multinomial expansion and use the fact that antenna gains, channel fading, and distances are independent across all the interferers, (c) comes from taking the expectation with respect to the fading random variable (one can find it by either considering series representation of the \ac{PDF} of fading or considering the integral in \cite{GradshteynRyzhik_2007}[Eq. 6.643.2]).

Now, the expectation in the expression above captures three (at least potentially) random parameters of any interfering transmitter, which are the blockage state, directionality and distance, and can be repesented as:
\begin{equation}
\begin{split}
    \mathbb{I}_1 &= \sum_{v\in\{\los,\nlos\}} p_{v}\sum_{g\in \Gc} p_{g}(\mu_v)_{k_i} \theta_{1,v}^{\mu_v}g^{k_i}\exp(-\theta_{2,v})\times\\
   &\Eb_R\Big[ \frac{l_v^{k_i}(R)}{\left(\frac{\zeta\theta_{1,t}}{\vartheta_{0,t}}g l_v(R) + \theta_{1,v} \right)^{k_i+\mu_v}}\confhypergeom{k_i+\mu_v}{\mu_v}{\frac{\theta_{1,v}\theta_{2,v}}{\frac{\zeta\theta_{1,t}}{\vartheta_{0,t}}g l_v(R) + \theta_{1,v}}}\Big],
   \end{split}
\end{equation}
where $p_g$ corresponds to the directionality gain probability model, and $p_{v}$ corresponds to the blockage probability model.

\section{Proof of \Corollary{ccdf_bpp_network_disc}}
\label{app:sinr_deterministic_network_disc}
We can find the expectation with respect to the distance of a user located in the center of a disk as follows:
\begin{equation}
    \begin{split}
    \Zf_{k_i} &= \Eb_R\left[ \frac{\left(l_v(r)\right)^{k_i}}{\left(\frac{\zeta\theta_{1,t}}{\vartheta_{0,t}}g l_v(r) + \theta_{1,v} \right)^{k_i+\mu_v}} \confhypergeom{k_i+\mu_v}{\mu_v}{\frac{\theta_{1,v}\theta_{2,v}}{\frac{\zeta\theta_{1,t}}{\vartheta_{0,t}}g l_v(r) + \theta_{1,v}}}\right]\\
     & = \frac{2}{\rad^2}\int_0^\rho \frac{\left(\gamma_v(r^2+\eta)^{-\alpha_v/2}\right)^{k_i}}{\left(\frac{\zeta\theta_{1,t}}{\vartheta_{0,t}}g \gamma_v(r^2+\eta)^{-\alpha_v/2} + \theta_{1,v} \right)^{k_i+\mu_v}} \confhypergeom{k_i+\mu_v}{\mu_v}{\frac{\theta_{1,v}\theta_{2,v}}{\frac{\zeta\theta_{1,t}}{\vartheta_{0,t}}g \gamma_v(r^2+\eta)^{-\alpha_v/2} + \theta_{1,v}}} r\dd{r}\\
     &\overset{(a)}{=} \frac{2\gamma_v^{k_i}}{\rad^2\theta_{1,v}^{k_i+\mu_v}}\int_h^{\sqrt{\rho^2+h^2}} \frac{y^{-\alpha_vk_i+1}}{\left(\frac{\zeta\theta_{1,t}}{\vartheta_{0,t}\theta_{1,v}}g \gamma_v y^{-\alpha_v} + 1 \right)^{k_i+\mu_v}} \confhypergeom{k_i+\mu_v}{\mu_v}{\frac{\theta_{2,v}}{\frac{\zeta\theta_{1,t}}{\vartheta_{0,t}\theta_{1,v}}g \gamma_v y^{-\alpha_v} + 1}} \dd{y}\\
     &= \frac{2\gamma_v^{k_i}}{\rad^2\theta_{1,v}^{k_i+\mu_v}}\sum_{j=0}^{\infty}\frac{(k_i+\mu_v)_j \theta_{2,v}^j}{(\mu_v)_j j!}
     \int_h^{\sqrt{\rho^2+h^2}} \frac{y^{-\alpha_vk_i+1}}{\left(\frac{\zeta\theta_{1,t}}{\vartheta_{0,t}\theta_{1,v}}g \gamma_v y^{-\alpha_v} + 1 \right)^{k_i+\mu_v+j}}\dd{y}\\
     &\overset{(b)}{=} \frac{2\gamma_v^{k_i}y^{2-k_i\alpha_v}}{\rad^2\theta_{1,v}^{k_i+\mu_v}(2-k_i\alpha_v)}\sum_{j=0}^{\infty}\frac{(k_i+\mu_v)_j \theta_{2,v}^j}{(\mu_v)_j j!} \gausshypergeom{k_i+\mu_v+j}{k_i-2/\alpha_v}{k_i-2/\alpha_v+1}{-\frac{\zeta\theta_{1,t}}{\vartheta_{0,t}\theta_{1,v}}g \gamma_vy^{-\alpha_v}}\Big|_h^{\sqrt{\rho^2+h^2}}\\
     &\overset{(c)}{=} \frac{2\gamma_v^{k_i}y^{2-k_i\alpha_v}}{\rad^2\theta_{1,v}^{k_i+\mu_v}(2-k_i\alpha_v)}\sum_{j=0}^{\infty}\sum_{l=0}^\infty\frac{(k_i+\mu_v)_j (k_i+\mu_v+j)_l (k_i-2/\alpha_v)_l}{(\mu_v)_j (k_i-2/\alpha_v+1)_l  j! l!} \theta_{2,v}^j\left(-\frac{\zeta\theta_{1,t}}{\vartheta_{0,t}\theta_{1,v}}g \gamma_vy^{-\alpha_v}\right)^l\Big|_h^{\sqrt{\rho^2+h^2}}\\
     &\overset{(d)}{=}\frac{2\gamma_v^{k_i}y^{2-k_i\alpha_v}}{\rad^2\theta_{1,v}^{k_i+\mu_v}(2-k_i\alpha_v)}\humbertpsione{k_i+\mu_v}{k_i-2/\alpha_v}{\mu_v}{k_i-2/\alpha_v+1}{\theta_{2,v}}{-\frac{\zeta\theta_{1,t}}{\vartheta_{0,t}\theta_{1,v}}g \gamma_vy^{-\alpha_v}}\Big|_h^{\sqrt{\rho^2+h^2}},\\
    \end{split}
\end{equation}
where for (a) we use variable transformation $y=\sqrt{r^2+\eta}$, (b) comes from \cite[3.194.1]{GradshteynRyzhik_2007} and is valid for $k_i-2/\alpha_v>0$, (c) uses the series representation of the hypergeometric function, (d) uses properties of the Humbert series $\Psi_1$.

\section{}
\label{app:list_symbols}
\begin{longtable}{p{.20\textwidth}p{.80\textwidth}}
\kill
\caption[]{Notation Used\label{tab:list_symbols}}
\\
\hline\hline
Symbol&Description\rule{0pt}{10pt}\\
\hline
$P$ & Pathloss (in dB) \\
$P_0$ & Pathloss at the reference distance (in dB) \\
$\alpha$ & Pathloss exponent \\
$d_0$ & Reference distance  \\
$d$ & Separation distance between the transmitter and receiver  \\
\multirow{2}[0]{*}{$\kappa$} & Ratio between  the total power in the dominant signal components \\
& and  the total power in the scattered signal components \\
$\mu$ & Number of multipath clusters\\
$\Omega$ & Mean signal power\\
$n_{\tx}$ & Number of access points\\
$\Phi$& Set of access points (point process)\\
$\Wc$ & Area considered\\
$\rho_0$ & Distance between the reference user and the origin\\
$\rad$ & Radius of the disk representing the considered area\\
$r_i$ & Distance to the $i$-th access point\\
$h_{\tx (\rx)}$ & Height of an access point (the reference user)\\
$\omega_{\tx (\rx)}$ & Transmitter (receiver) beamwidth\\ 
$\gainM_{\tx (\rx)}$ & Transmitter (receiver) mainlobe gain\\
$\gainm_{\tx (\rx)}$ & Transmitter (receiver) sidelobe gain\\
$G_i$ & Alignment gain with the $i$-th access point\\
$p_{g_i}$ & Probability mass function of the event $G_i=g_i$\\
$p_{\nlos}$ & Blockage probability \\
\multirow{2}[0]{*}{$\gamma_{i,t}$} & Pathloss at the reference distance in linear scale given access point $i$ and\\
& blockage state $t$\\
$\alpha_t$ & Pathloss exponent given blockage state $t$\\
$H_{i,t}$ & Blockage-dependent power fading\\
$\{\kappa_t,\mu_t,\Omega_t\}$ & Parameters of the fading model given blockage state $t$\\
$l_{t}\left(\cdot\right)$ & Blockage-dependent pathloss\\ 
$S_{i,t}$ & Received power from the $i$-th access point given blockage state $t$\\
$I$ & Interference power\\
$\tau$ & \acl{SNR}\\
$\zeta$ & \acl{SINR} threshold\\
$F^c_{X}(\cdot)$ &\acl{CCDF} of $X$\\
$C_{\mathrm{area}}$ & \acl{ATC}\\
$\lambda=\frac{n_{\tx}}{|\Wc|}$ & \acl{AP} density\\
$\mathrm{bw}$ & System bandwidth\\
$\mathrm{SE}$ & Spectral efficiency\\
$q_X(\beta)$ & $\beta$-quantile of $X$\\
$f_X(\cdot)$ & \acl{PDF} of $X$\\
$\mathcal{N}$ & Set of candidate locations for deploying \acl{mmWave} access points\\
$K$ & Number of annuli constituting the floor of the considered area \\
$M_i$ & Number of circles constituting the $i$-th annulus \\
$r_d$ & Radius of the geographical area of interest \\
$r_b$ & Radius of each circular area in the geographical area of interest \\ 
$\mathcal{K}$ & Set of circular areas in the floor of the considered area \\
$A_k$ &$k$-th circular area in $\mathcal{K}$\\
$p_{nk}$ & Probability that the link between an \acl{AP} at location $n$ and $A_k$ is available \\
$B$ & Maximum number of beams that an \acl{AP} can have \\
$\beta$ & Requested coverage probability \\
$\Pb_{\text{cov}}$ & Network-wide coverage probability \\
\hline
\end{longtable}

\section*{Acknowledgment}

This material is based in part upon work supported by the National Science Foundation under grant CNS-1526844, the Science Foundation Ireland under grant 14/US/I3110 and the Department for the Economy Northern Ireland through grant USI080. The authors would also like to thank Shubhajeet Chatterjee and Fadhil Firyaguna, for their support at various stages of the preparation of this manuscript.

\ifCLASSOPTIONcaptionsoff
  \newpage
\fi



%

\bibliographystyle{./bibtex/IEEEtran}
\bibliography{./bibtex/IEEEabrv,./bibtex/literature_proposal,./bibtex/deployment,./bibtex/channel,./bibtex/standardization}


\end{document}